\documentclass[10pt,epsf,preprint]{aastex}
\usepackage{epsf}
\input psfig 

\begin{document}
\title{Early-type galaxies in the SDSS. IV. Colors and chemical evolution}

\author{
Mariangela Bernardi\altaffilmark{\ref{Chicago},\ref{CMU}},
Ravi K. Sheth\altaffilmark{\ref{Fermilab},\ref{Pitt}},
James Annis\altaffilmark{\ref{Fermilab}},
Scott Burles\altaffilmark{\ref{Fermilab}},
Douglas P. Finkbeiner\altaffilmark{\ref{Berkeley},\ref{Princeton},\ref{HF}},
Robert H. Lupton\altaffilmark{\ref{Princeton}},
David J. Schlegel\altaffilmark{\ref{Princeton}}, 
Mark Subbarao\altaffilmark{\ref{Chicago}},
Neta A. Bahcall\altaffilmark{\ref{Princeton}},
John P. Blakeslee\altaffilmark{\ref{JHU}},
J. Brinkmann\altaffilmark{\ref{APO}},
Francisco J. Castander\altaffilmark{\ref{yale},\ref{chile}},
Andrew J. Connolly\altaffilmark{\ref{Pitt}}, 
Istvan Csabai\altaffilmark{\ref{Eotvos},\ref{JHU}},
Mamoru Doi\altaffilmark{\ref{Tokyo1},\ref{Tokyo2}},
Masataka Fukugita\altaffilmark{\ref{ICRR},\ref{IAS}},
Joshua Frieman\altaffilmark{\ref{Chicago},\ref{Fermilab}},
Timothy Heckman\altaffilmark{\ref{JHU}},
Gregory S. Hennessy\altaffilmark{\ref{USNO}},
\v{Z}eljko Ivezi\'{c}\altaffilmark{\ref{Princeton}},
G. R. Knapp\altaffilmark{\ref{Princeton}},
Don Q. Lamb\altaffilmark{\ref{Chicago}},
Timothy McKay\altaffilmark{\ref{UMich}},
Jeffrey A. Munn\altaffilmark{\ref{USNO}},
Robert Nichol\altaffilmark{\ref{CMU}},
Sadanori Okamura\altaffilmark{\ref{Tokyo3},\ref{Tokyo2}}, 
Donald P. Schneider\altaffilmark{\ref{PSU}},
Aniruddha R. Thakar\altaffilmark{\ref{JHU}},
and Donald G.\ York\altaffilmark{\ref{Chicago}}
}

\newcounter{address}
\setcounter{address}{1}
\altaffiltext{\theaddress}{\stepcounter{address}
University of Chicago, Astronomy \& Astrophysics
Center, 5640 S. Ellis Ave., Chicago, IL 60637\label{Chicago}}
\altaffiltext{\theaddress}{\stepcounter{address}
Department of Physics, Carnegie Mellon University, Pittsburgh, PA 15213
\label{CMU}}
\altaffiltext{\theaddress}{\stepcounter{address}
Fermi National Accelerator Laboratory, P.O. Box 500,
Batavia, IL 60510\label{Fermilab}}
\altaffiltext{\theaddress}{\stepcounter{address}
Department of Physics and Astronomy, University of Pittsburgh, Pittsburgh, PA 15620\label{Pitt}}
\altaffiltext{\theaddress}{\stepcounter{address}
Department of Astronomy, University of California at Berkeley, 601 Campbell Hall, Berkeley, CA 94720\label{Berkeley}}
\altaffiltext{\theaddress}{\stepcounter{address}
Princeton University Observatory, Princeton, NJ 08544\label{Princeton}}
\altaffiltext{\theaddress}{\stepcounter{address}Hubble Fellow\label{HF}}
\altaffiltext{\theaddress}{\stepcounter{address}
Department of Physics \& Astronomy, The Johns Hopkins University, 3400 North Charles Street, Baltimore, MD 21218-2686\label{JHU}}
\altaffiltext{\theaddress}{\stepcounter{address}
Apache Point Observatory, 2001 Apache Point Road, P.O. Box 59, Sunspot, NM
88349-0059\label{APO}}
\altaffiltext{\theaddress}{\stepcounter{address} Yale University, P. O. Box
208101, New Haven, CT 06520\label{yale}}
\altaffiltext{\theaddress}{\stepcounter{address} Universidad de Chile, Casilla
36-D, Santiago, Chile\label{chile}}
\altaffiltext{\theaddress}{\stepcounter{address}
Department of Physics of Complex Systems, E\"otv\"os University, Budapest, H-1117 Hungary\label{Eotvos}}
\altaffiltext{\theaddress}{\stepcounter{address}
Institute of Astronomy, School of Science, University of Tokyo, Mitaka, Tokyo 181-0015, Japan\label{Tokyo1}}
\altaffiltext{\theaddress}{\stepcounter{address}
Research Center for the Early Universe, School of Science,
    University of Tokyo, Tokyo 113-0033, Japan\label{Tokyo2}}
\altaffiltext{\theaddress}{\stepcounter{address}
Institute for Cosmic Ray Research, University of Tokyo, Kashiwa 277-8582, Japan\label{ICRR}}
\altaffiltext{\theaddress}{\stepcounter{address}
Institute for Advanced Study, Olden Lane, Princeton, NJ 08540\label{IAS}}
\altaffiltext{\theaddress}{\stepcounter{address}
U.S. Naval Observatory, 3450 Massachusetts Ave., NW, Washington, DC 20392-5420\label{USNO}}
\altaffiltext{\theaddress}{\stepcounter{address}
Department of Physics, University of Michigan, 500 East University, Ann Arbor, MI 48109\label{UMich}}
\altaffiltext{\theaddress}{\stepcounter{address}
Department of Astronomy, University of Tokyo,
   Tokyo 113-0033, Japan\label{Tokyo3}}
\altaffiltext{\theaddress}{\stepcounter{address}
Department of Astronomy and Astrophysics, The Pennsylvania State University, University Park, PA 16802\label{PSU}}



\begin{abstract}
The colors and chemical abundances of early-type galaxies at redshifts 
$z<0.3$ are studied using a sample of nearly 9000 galaxies, selected 
from the Sloan Digital Sky Survey using morphological and spectral 
criteria.  
In this sample, redder galaxies have larger velocity dispersions:  
$g^*-r^*\propto\sigma^{0.26\pm 0.02}$.  Color also correlates with 
magnitude, $g^*-r^*\propto (-0.025\pm 0.003)\,M_{r_*}$, and size, 
but these correlations are entirely due to the $L-\sigma$ and 
$R_o-\sigma$ relations:  the primary correlation is color$-\sigma$.  
The red light in early-type galaxies is, on average, slightly more 
centrally concentrated than the blue.  Because of these color gradients, 
the strength of the color--magnitude relation depends on whether or 
not the colors are defined using a fixed metric aperture; 
the color$-\sigma$ relation is less sensitive to this choice.  

Chemical evolution and star formation histories of early-type galaxies 
are investigated using co-added spectra of similar objects.  
The resulting library of co-added spectra contains spectra that 
represent a wide range of early-type galaxies.
Chemical abundances correlate primarily with velocity dispersion:  
H$_\beta\propto\sigma^{-0.24\pm 0.03}$, 
Mg$_2\propto\sigma^{0.20\pm 0.02}$, Mg$b\propto\sigma^{0.32\pm 0.03}$, 
and $\langle{\rm Fe}\rangle\propto\sigma^{0.11\pm 0.03}$.  
At fixed $\sigma$, the population at $z\sim 0.2$ had weaker Mg$_2$ 
and stronger H$_\beta$ absorption compared to the population at 
$z\sim 0$.  It was also bluer.  
Comparison of these colors and line-strengths, and their evolution, 
with single-burst stellar population models suggests a formation 
time of 9~Gyrs ago, consistent with a Fundamental Plane analysis 
of this sample.  

Although the Fundamental Plane shows that galaxies in dense regions 
are slightly different from galaxies in less dense regions, the 
co-added spectra and color--magnitude relations show no statistically 
significant dependence on environment.  
\end{abstract}  

\section{Introduction}
This is the fourth in a series of papers which studies the properties 
of early-type galaxies at relatively low redshift $z\le 0.3$.   
Paper~I (Bernardi et al. 2003a) describes how we extracted the sample 
from the SDSS database, 
and how the photometric and spectroscopic parameters (luminosities, 
effective radii, surface brightnesses, colors and velocity dispersions) 
were estimated.  It also provides the tables of these parameters.  
Paper~II (Bernardi et al. 2003b) studies the luminosity function, and 
various early-type galaxy correlations in multiple bands 
($g^*$, $r^*$, $i^*$ and $z^*$). 
Paper~III (Bernardi et al. 2003c) studies the Fundamental Plane, 
and its dependence on wavelength, redshift and environment.  
In this fourth paper, we study the colors and the spectral line 
indices of the galaxies in our sample, both of which correlate 
strongly with velocity dispersion $\sigma$.    

Section~\ref{cms} presents color--magnitude and color--velocity 
dispersion relations.  It shows that the primary correlation is 
color--$\sigma$; color--size and color--magnitude relations are 
a consequence of the fact that size and magnitude correlate with 
$\sigma$.  
In Section~\ref{lindices}, the spectra of the galaxies in our 
sample are used to study if and how the chemical composition of the 
early-type galaxy population depends on redshift and environment.  
The signal-to-noise ratios of the spectra in this SDSS sample are 
substantially smaller than the $S/N=100$ required to estimate the 
Lick indices reliably.  Therefore, Section~\ref{coadd} describes 
the procedure we have adopted for obtaining reliable estimates of 
absorption line strengths and presents the line indices measurements 
we used in our analysis. One of the results of this paper is a 
library of co-added spectra which contains spectra that represent 
a wide range of early-type galaxies.  This library is available 
electronically. Sections~\ref{fixedv} and~\ref{lindexc} 
show correlations with velocity dispersion and color respectively. 
Section~\ref{ssp} compares these measurements with predictions from 
single burst stellar population models, and Section~\ref{environ} 
studies how these trends depend on environment.  

We have chosen to present results for Mg$_2$ (measured in magnitudes), 
and Mg$b$, $\langle{\rm Fe}\rangle$ and H$_\beta$ (measured in Angstroms), 
where $\langle{\rm Fe}\rangle$ represents an average over Fe5270 and Fe5335.  
Mg$_2$ and Mg$b$ are alpha elements, so, roughly speaking, they reflect 
the occurence of Type II supernovae, whereas Fe is produced in SN~Ia.  
All these line indices depend both on the age and the metallicity of 
the stellar population (e.g., Worthey 1994), although Mg and Fe are 
more closely related to the metallicity, whereas the equivalent width 
of H$_\beta$ is an indicator of recent star formation.  
An analysis of other indices on a similar set of co-added SDSS spectra is 
presented in Eisenstein et al. (2003).

Except where stated otherwise, we write the Hubble constant as 
$H_0=100\,h~\mathrm{km\,s^{-1}\,Mpc^{-1}}$, and we perform our analysis 
in a cosmological world model with 
$(\Omega_{\rm M},\Omega_{\Lambda},h)=(0.3,0.7,0.7)$, where 
$\Omega_{\rm M}$ and $\Omega_{\Lambda}$ are the present-day scaled 
densities of matter and cosmological constant.  
In such a model, the age of the Universe at the present time is 
$t_0=9.43h^{-1}$ Gyr.  

\section{The color--magnitude and color$-\sigma$ relations}\label{cms}

The colors of early-type galaxies are observed to correlate with 
their luminosities, with small scatter around the mean relation 
(e.g., Baum 1959; de Vaucouleurs 1961; Sandage \& Visvanathan 1978a,b; 
Bower, Lucey \& Ellis 1992a,b).  In this section we examine these 
correlations using the model colors output by the SDSS photometric 
pipeline.  Section~\ref{cz} shows that the color--magnitude relation 
in our sample is evolving:  the population at higher redshift is 
bluer.  It also shows that the primary correlation is actually color 
with velocity dispersion:  color--magnitude and color--size relations 
arise simply because magnitude and size are also correlated with 
velocity dispersion.  Section~\ref{cden} shows that the color--velocity 
dispersion relation exhibits no significant dependence on environment.  
It has been known for some time that giant early-type galaxies are 
reddest in their cores and become bluer toward their edges (e.g., 
de Vaucouleurs 1961; Sandage \& Visvanathan 1978a).  
Therefore, the strength the color-magnitude relation depends on how 
the color is defined.  This is the subject of Section~\ref{cgrad}.

\subsection{Galaxy colors:  evolution}\label{cz}

\begin{figure}
\centering
\epsfxsize=\hsize\epsffile{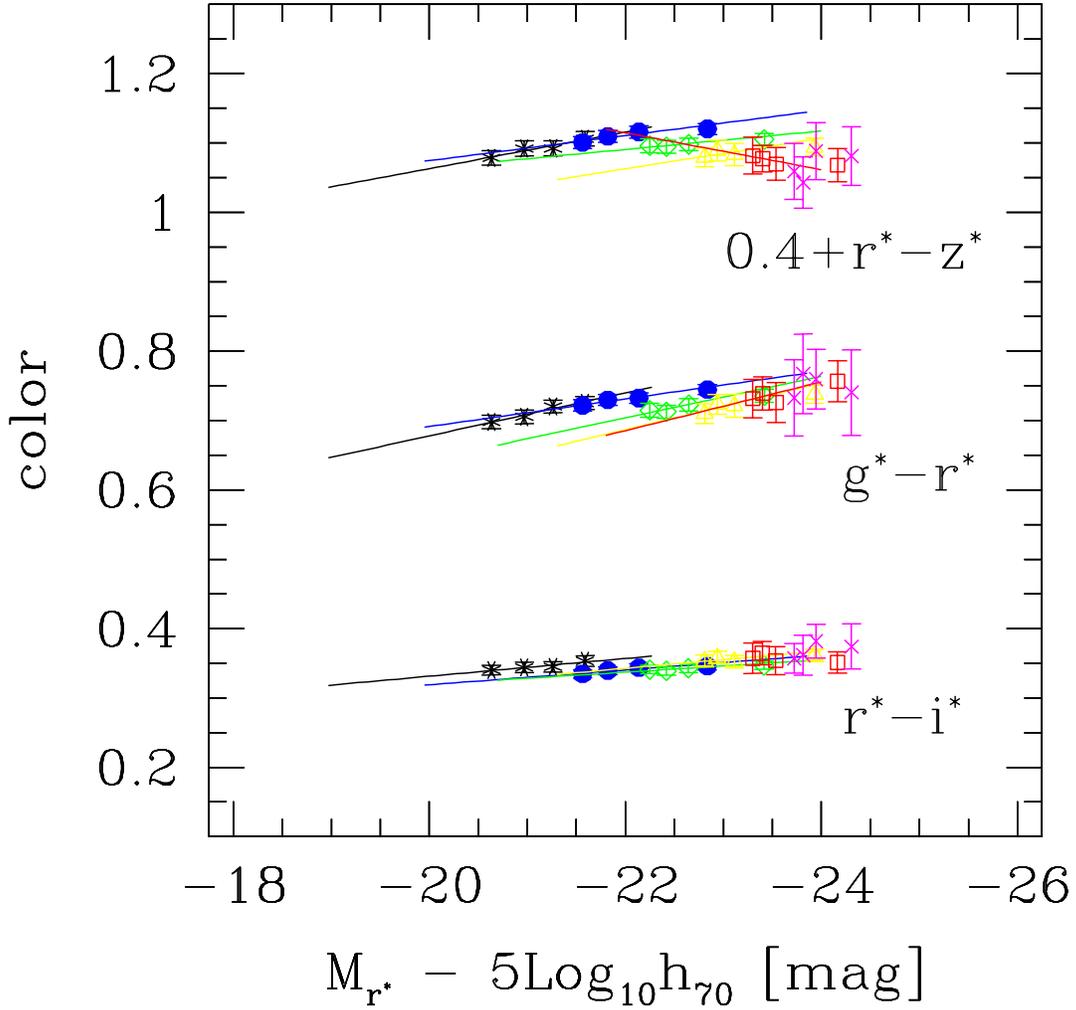}
\caption[]{Color versus $r^*$ magnitude in volume limited subsamples.  
Symbols show the median color at fixed luminosity as measured in the 
different volume limited subsamples, error bars show three times the 
uncertainty in this median.  Dashed lines show linear fits to the relation 
in each subsample.  The slope of the relation is approximately the same in 
all the subsamples, although the relations in the more distant subsamples 
are offset blueward.  This offset is greater for the $g^*-i^*$ colors than 
for $r^*-i^*$.  Because of this offset, the slope of a line which passes 
through the relation defined by the whole sample is very different from 
the slope in each of the subsamples.}
\label{cmag3}
\end{figure}

\begin{figure}
\centering
\epsfxsize=\hsize\epsffile{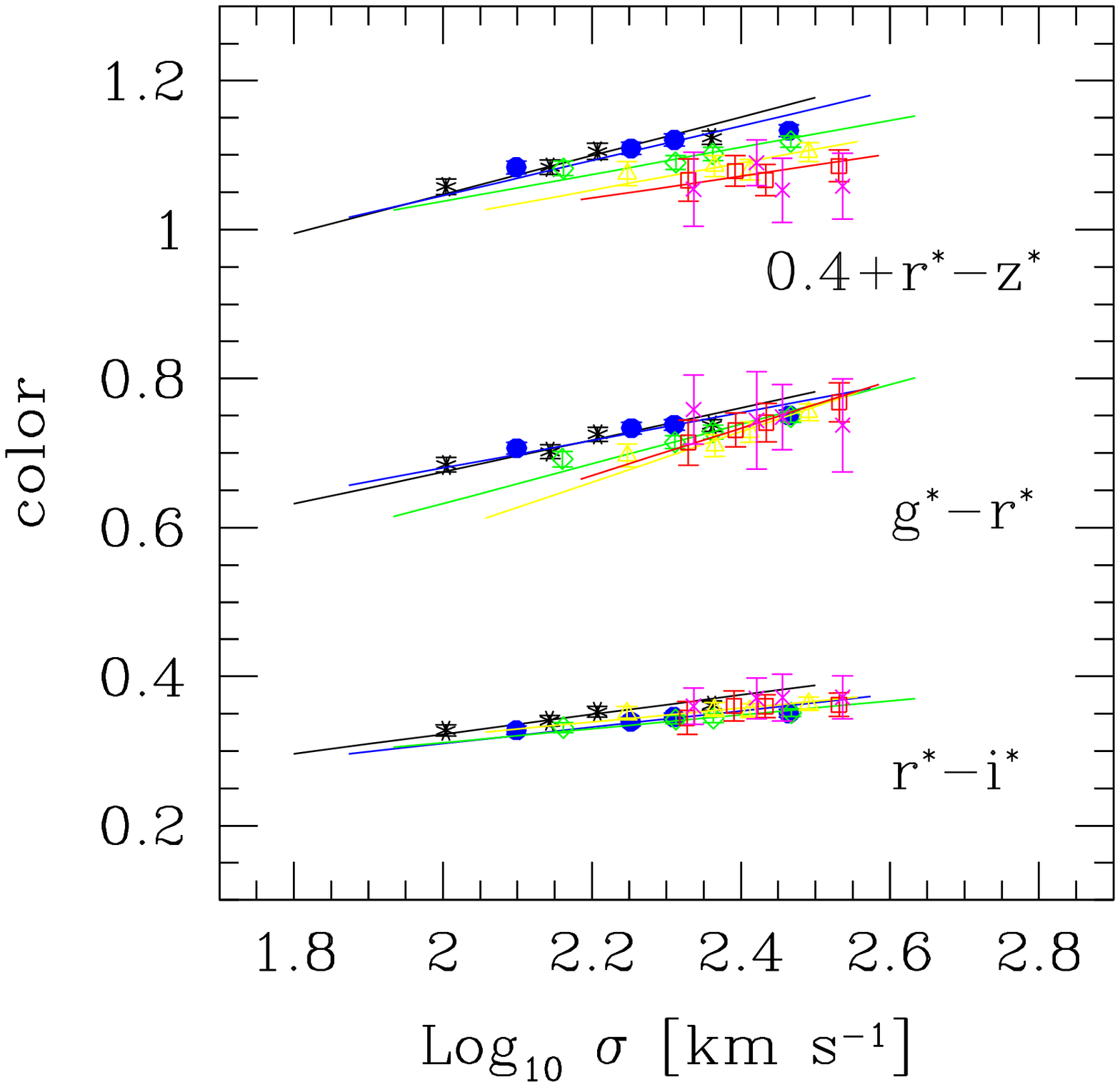}
\caption[]{Same as previous figure, but now showing color versus 
velocity dispersion.  Redder galaxies have larger velocity dispersions.  
Dashed lines show that the slope of the relation is approximately the 
same in all the subsamples, but that the relations in the more distant 
subsamples are offset blueward.  The offset is similar to that in the 
color--magnitude relations.}
\label{csig3}
\end{figure}

We begin with a study of the color--magnitude relation in our data set.  
Estimating the slope of this relation is complicated because our sample 
is magnitude limited and spans a relatively wide range of redshifts, and 
because the slope of the color--magnitude relation is extremely shallow.  
At any given redshift, we do not have a wide range of magnitudes over 
which to measure the relation.  If we are willing to assume that this 
relation does not evolve, then the different redshift bins probe different 
magnitudes, and we can build a composite relation by stacking together 
the relations measured in any individual redshift bin.  
However, the shallow slope of the relation means that small changes in 
color, whether due to measurement errors or evolution, result in large 
changes in $M$.  Thus, if the colors of early-type galaxies evolve even 
weakly, the slope of the composite color--magnitude relation is drastically 
affected.  We can turn this statement around, of course, and use the 
color--magnitude relation as a sensitive test of whether or not the colors 
of the galaxies in our sample have evolved.  

Figure~\ref{cmag3}, the relation between the absolute magnitude in 
$r^*$ and the $g^*-r^*$, $r^*-i^*$ and $r^*-z^*$ colors, illustrates 
our argument.  We have chosen to present results for these three 
colors only, since the other colors in our dataset are just linear 
combinations of these, and because, as described in Paper~I, the 
$r^*$ band plays a special role in the SDSS photometry.  Briefly, 
this is the band in which the SDSS spectroscopic sample is selected, 
and this band has a special status with regard to the SDSS `model' 
colors (c.f. Section~\ref{cgrad}).  

The figure was constructed by using the same volume limited 
subsamples we used when analyzing the $r^*$ luminosity function in 
Paper~II.  
Symbols with error bars show the median, and the error in this median, 
at fixed luminosity in each subsample.  Dashed lines show the mean color 
at fixed magnitude in each subsample; the slopes of these mean relations 
and the scatter around the mean are approximately the same (we will 
quantify the slopes of these relations shortly) but the zero-points 
are significantly different.  
All three color--magnitude relations show qualitatively similar trends, 
namely a tendency to shift blueward with increasing redshift.  
For example, $r^*-i^*$ is bluer by about 0.03~mags in the most distant 
subsample than in the nearest, whereas the shift in $g^*-i^*$ is 
closer to 0.09 mags.  
Because of the blueward shifts, the slope of a linear fit to the whole 
catalog, over the entire range in absolute magnitudes shown, is much 
shallower than the slopes of the individual subsamples.

How much of the evolution in Figure~\ref{cmag3} is due to changes in 
color, and how much to changes in luminosity?  
To address this, Figure~\ref{csig3} shows the same plot, but with $r^*$ 
magnitude replaced by velocity dispersion.  As before, the different 
dashed lines show fits to the color-$\sigma$ relations in the individual 
subsamples; the slopes of, and scatter around, the mean relations are 
similar but the zero-points are different.  The magnitude of the shift 
in color is similar to what we found for the color--magnitude relation, 
suggesting that the offsets are due primarily to changes in colors rather 
than luminosity.  

At first sight this might seem surprising, because single-burst models 
suggest that the evolution in the colors is about one-third that of 
the luminosities.  However, because the slope of the color--magnitude 
relation is so shallow, even a large change in magnitudes produces only 
a small shift in the zero-point of the colors.  To illustrate, let 
$(C - C_*) = -0.02(M - M_*)$ denote the color--magnitude relation at 
the present time.  Now let the typical color and magnitude change by 
setting $C_*\to C_* + \delta C$ and $M_*\to M_* + \delta M$, but assume 
that the slope of the color--magnitude relation does not.  This 
corresponds to a shift in the zero-point of $0.02\delta M + \delta C$,  
demonstrating that $\delta C$ dominates the change in the zero point 
even if it is a factor of ten smaller than $\delta M$. (Note that 
a shallow color--magnitude relation was also obtained with total 
optical-to-near infrared colors by Fioc \& Rocca-Volmerange 1999). 

\begin{table}[t]
\centering
\caption[]{Maximum-likelihood estimates of the joint distribution of 
color, $r^*$ magnitude and velocity dispersion and its evolution.  
At redshift $z$, the mean values are $C_*-Pz$, $M_*-Qz$, and $V_*$, 
and the covariances are $\langle(C-C_*)(M-M_*)\rangle = 
              \sigma^2_{CM}=\sigma_C\sigma_M\,\rho_{CM}$ etc.\\}
\begin{tabular}{cccccccccccc}
\tableline 
Color & $C_*$ & $\sigma_C$ & $V_*$ & $\sigma_V$ & $M_*$ 
& $\sigma_M$ & $\rho_{CM}$ & $\rho_{CV}$ & $\rho_{VM}$ & Q & P \\
\hline\\
$g^*-r^*$ & 0.736 & 0.0570 & 2.200 & 0.1112 & $-21.15$ & 0.841 & $-0.361$ &
0.516 & $-0.774$ & 0.85 & 0.30\\
$r^*-i^*$ & 0.346 & 0.0345 & 2.200 & 0.1112 & $-21.15$ & 0.841 & $-0.301$ &
0.401 & $-0.774$ & 0.85 & 0.10\\
$r^*-z^*$ & 0.697 & 0.0517 & 2.203 & 0.1114 & $-21.15$ & 0.861 & $-0.200$ &
0.346 & $-0.774$ & 0.85 & 0.15\\
\tableline
\end{tabular}
\label{MLcmag}
\end{table}

\begin{figure}
\centering
\epsfxsize=\hsize\epsffile{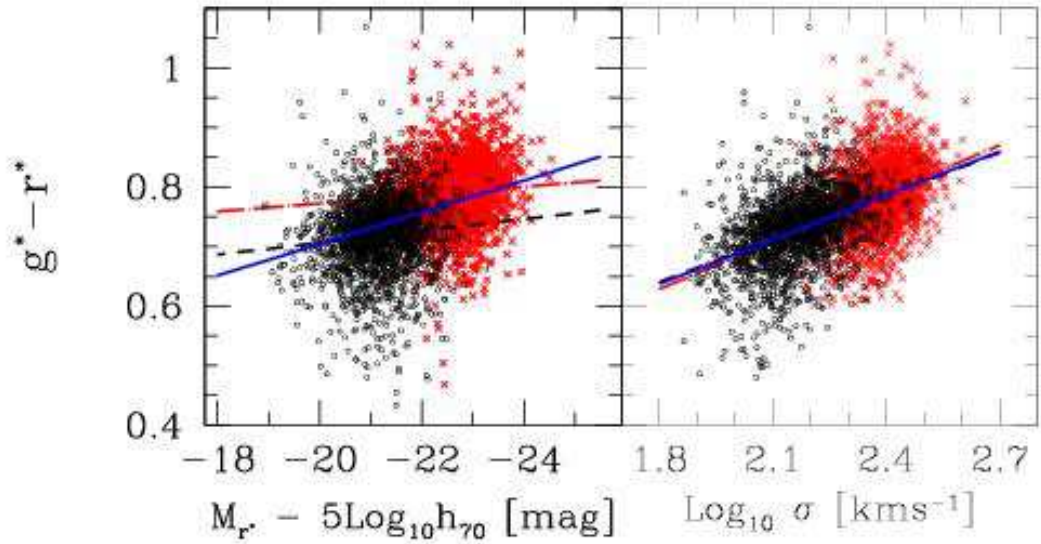}
\vspace{-5cm}
\caption[]{Relation between color and magnitude at fixed velocity 
dispersion (left) and between color and velocity dispersion at fixed 
magnitude (right).  In the panel on the left the correlation between 
color and magnitude is much weaker in the two subsamples than it is 
for the whole sample, indicating that the color--magnitude relation 
is driven by the dependence of color and magnitude on velocity 
dispersion.  On the other hand, in the panel on the right, the 
individual fits to the two subsamples are indistinguishable from 
the fits to the whole sample, indicating that the correlation between 
color and $\sigma$ does not depend on magnitude.   }
\label{clrmagv}
\end{figure}

To account both for selection effects and evolution, we have computed 
maximum likelihood estimates of the joint color--magnitude--velocity 
dispersion distribution, allowing for evolution in the magnitudes and 
the colors but not in the velocity dispersions:  
i.e., the magnitudes and colors are assumed to follow Gaussian 
distributions around mean values which evolve, say $M_*(z) = M_* - Qz$ 
and $C_*(z) = C_* - Pz$, but the spread around the mean values, and the 
correlations between $C$ and $M$ do not evolve.  (The maximum 
likelihood technique we use is described in more detail in Paper~II.)
We have chosen to only present results for the color--$r^*$-magnitude 
relation, because, as we argue later, color correlates primarily with 
$\sigma$, which is independent of waveband.  

The results are summarized in Table~\ref{MLcmag}.  
Notice that the colors at redshift zero are close to those of the 
Coleman, Wu \& Weedman (1980) templates; that the evolution in color 
is smaller than in magnitude, and consistent with the individual 
estimates of the evolution in the different bands (Table~1 of Paper~II 
shows that $Q=1.15, 0.75$ and 0.60 in the $g^*$, $i^*$ and $z^*$ 
bands respectively); 
and that the best fit distributions of $M_*$ and $V_*$ are the same for 
all three colors, and are similar to the values we found in Paper~II.  

As discussed in Paper~II and Paper~III, various combinations of the 
coefficients in Table~\ref{MLcmag} yield maximum likelihood estimates 
of the slopes of linear regressions of pairs of variables.  Some of 
these are summarized in Table~\ref{CMVslopes}.  One interesting 
combination is the relation between color and magnitude at fixed 
velocity dispersion:  
\begin{displaymath}
{\Bigl\langle C- \langle C|V\rangle \big|M\Bigr\rangle\over\sigma_{C|V}} = 
{M-\langle M|V\rangle\over \sigma_{M|V}} \times 
{(\rho_{CM} - \rho_{CV}\rho_{VM})\over\sqrt{(1-\rho_{VM}^2)(1-\rho_{CV}^2)}}
\qquad {\rm at\ fixed\ } V=\log_{10}\sigma.
\end{displaymath}
Inserting the values from Table~\ref{MLcmag} shows that, at fixed 
velocity dispersion, there is little correlation between color and 
luminosity.  In other words, the color--magnitude relation is almost 
entirely due to the correlation between color and velocity dispersion.  

Figure~\ref{clrmagv} shows this explicitly.  The dashed and dot--dashed 
lines show fits to the relation between color and magnitude at low 
(circles) and high (crosses) velocity dispersion (in the plots,
the maximum likelihood estimates of the evolution in color and magnitude 
have been removed).  The solid line shows the color--magnitude relation 
for the full sample which includes the entire range of $\sigma$; it is 
considerably steeper than the relation in either of the subsamples.  
The panel on the right shows the color$-\sigma$ relation at low (circles) 
and high (crosses) luminosity.  The individual fits to the two 
subsamples are indistinguishable from the fits to the whole sample.  

This is also true for the color--size relation, although we have not 
included a figure showing this.  One consequence of this is that 
residuals from the Faber--Jackson relation correlate with color, 
whereas residuals from the luminosity--size relation do not.  We will 
return to this later.  Because the primary correlation is color with 
velocity dispersion, in what follows, we will mainly consider the 
color--$\sigma$ relation, and residuals from it.  

\begin{table}[t]
\centering
\caption[]{Maximum-likelihood estimates of the slopes and zero-points 
of the color-at-fixed-magnitude and color-at-fixed-velocity dispersion 
relations, and the scatter around the mean relations.  \\}
\begin{tabular}{ccccccc}
\tableline 
Color & slope & zero-point & rms & slope & zero-point & rms \\
\hline
 &\multicolumn{3}{c}{color$-r^*$ magnitude} & 
  \multicolumn{3}{c}{color$-\log_{10}\sigma$} \\
$g^*-r^*$ & $-0.025\pm 0.003$ & 0.218 & 0.053 & $0.26\pm 0.02$ & 0.154 & 0.0488 \\
$r^*-i^*$ & $-0.012\pm 0.002$ & 0.085 & 0.033 & $0.12\pm 0.02$ & 0.072 & 0.0316 \\ 
$r^*-z^*$ & $-0.012\pm 0.003$ & 0.443 & 0.051 & $0.16\pm 0.02$ & 0.343 & 0.0485 \\
\tableline
\end{tabular}
\label{CMVslopes}
\end{table}

While the color--$\sigma$ provides clear evidence that the colors in 
the high redshift population in our sample are bluer than in the nearby 
population, quantifying how much the colors have evolved is more 
difficult, because the exact amount of evolution depends on the 
K-correction we assume.  Appendix~A of Paper~I discusses how we make 
our K-corrections, as well as what this choice implies for our 
estimated color evolution.  

\subsection{Galaxy colors:  environment}\label{cden}

\begin{figure}
\centering
\epsfxsize=1.1\hsize\epsffile{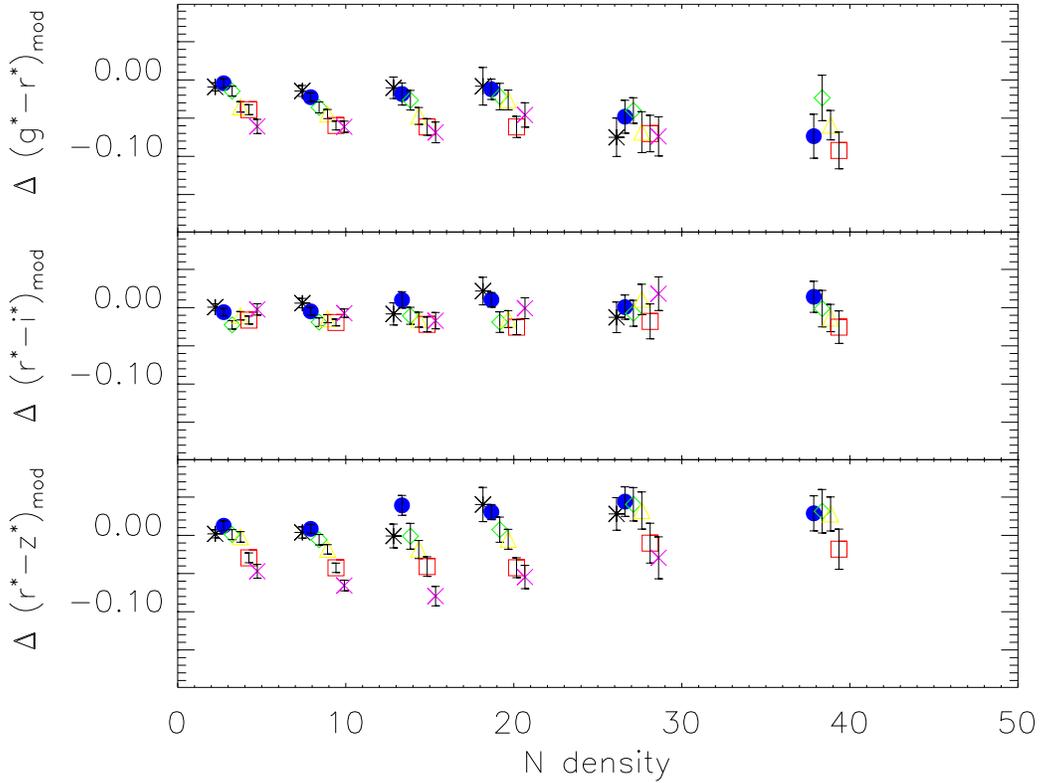}
\vspace{-3cm}
\caption[]{Residuals from the color-$\sigma$ relation as a function 
of local density.  At each bin in density, symbols showing results 
for higher redshifts have been offset slightly to the right.  
Galaxies at higher redshifts are bluer---hence the trend to slope 
down and to the right at fixed $N$.  The $(r^*-z^*)$ colors of galaxies 
in dense environments are redder than those of their counterparts in 
less dense regions, although the trend is weaker in the other colors.  
Although the $g^*-r^*$ color appears to show the opposite trend, note 
that the lowest redshift densest bin is the one in which our grouping 
algorithm is least secure.}  
\label{cmdensity}
\end{figure}
\begin{figure}
\centering
\epsfxsize=1.1\hsize\epsffile{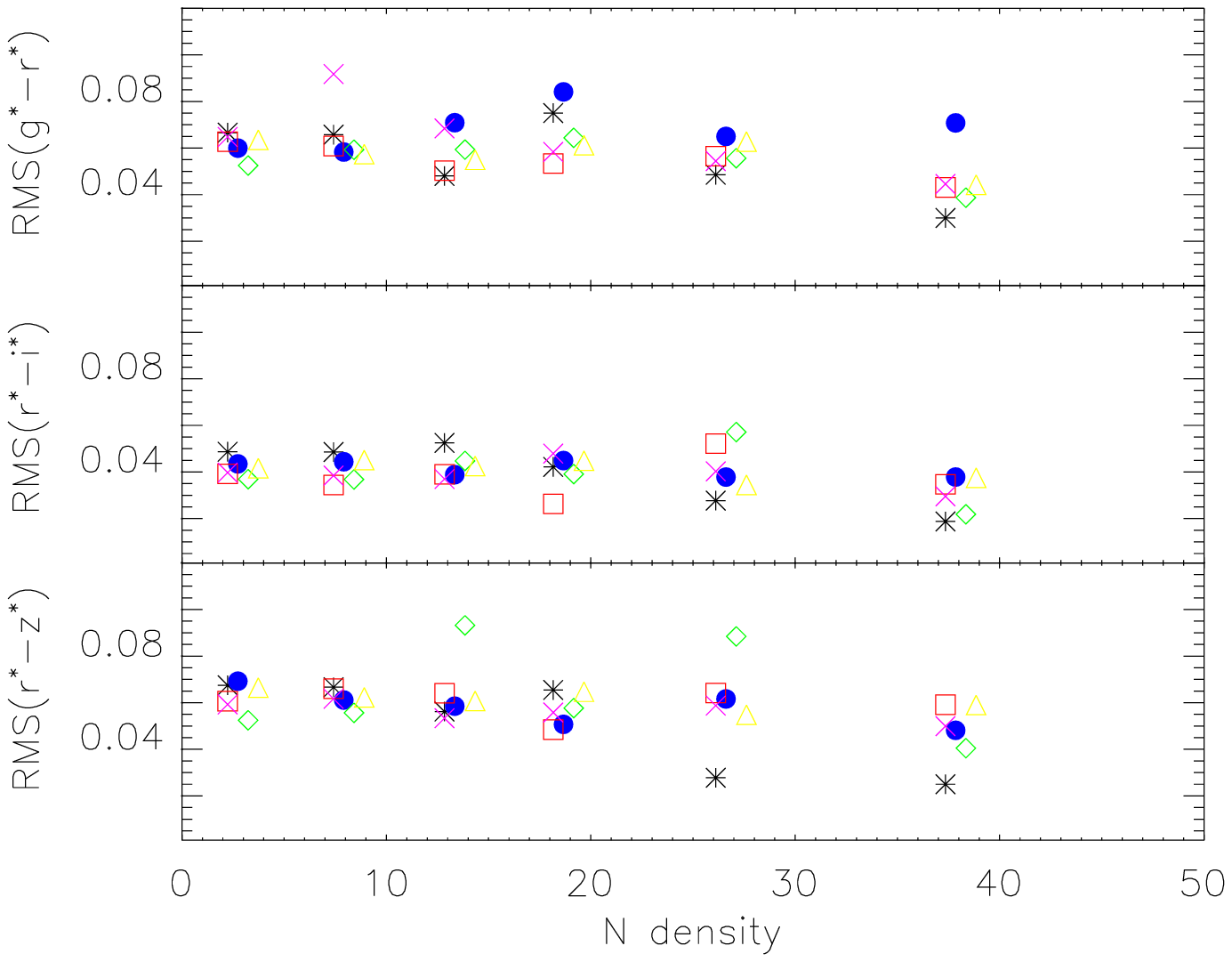}
\vspace{-3cm}
\caption[]{As for the previous figure, but now showing the thickness 
of the color-$\sigma$ relation as a function of local density.  }
\label{rmsdensity}
\end{figure}

Having shown that the colors are evolving, and that, to a reasonable 
approximation, this evolution affects the amplitude but not the slope 
of the color--magnitude and color--$\sigma$ relations, we now study 
how the colors depend on environment.  
To present our results, we assume that environmental effects affect 
the amplitude more strongly than the slope of the color--$\sigma$ 
relation.  Therefore, we assume that the slope is fixed, and fit for 
the shift in color which best describes the subsample.  
Figure~\ref{cmdensity} shows the results.  As in Papers~~I, II and~III, 
galaxies were divided into different bins in local density, and then 
further subdivided by redshift (the local density was estimated using 
the number of near neighbours in coordinate and color space, see 
Paper~I for details).  Different symbols in each bin in local density 
show results for the different redshifts; higher redshifts are offset 
slightly to the right.  This makes trends with evolution easy to 
separate from those due to environment.  In addition to the 
evolutionary trends we have just discussed, the figure shows that 
the $r^*-z^*$ colors are redder in denser regions (bottom panel), 
but that this trend is almost completely absent for the other colors.  
The $g^*-r^*$ color appears to be slightly bluer in the densest 
region lowest redshift bin, where our grouping procedure is least 
secure.  

The tightness of the colour-magnitude relation of cluster early-types  
has been used to put constraints on the ages of cluster early-types 
(e.g., Kodama et al. 1999).  Figure~\ref{rmsdensity} shows how the 
thickness of this relation depends on environment.  The plot shows 
no evidence that the scatter around the mean relation decreases 
slightly with increasing density; a larger sample is needed to make 
conclusive quantitative statements about this, and about whether or 
not the scatter around the mean relation depends more strongly on 
environment at low than at high redshift.

\subsection{Color gradients and the color--magnitude relation}\label{cgrad}
\begin{figure}
\centering
\epsfxsize=\hsize\epsffile{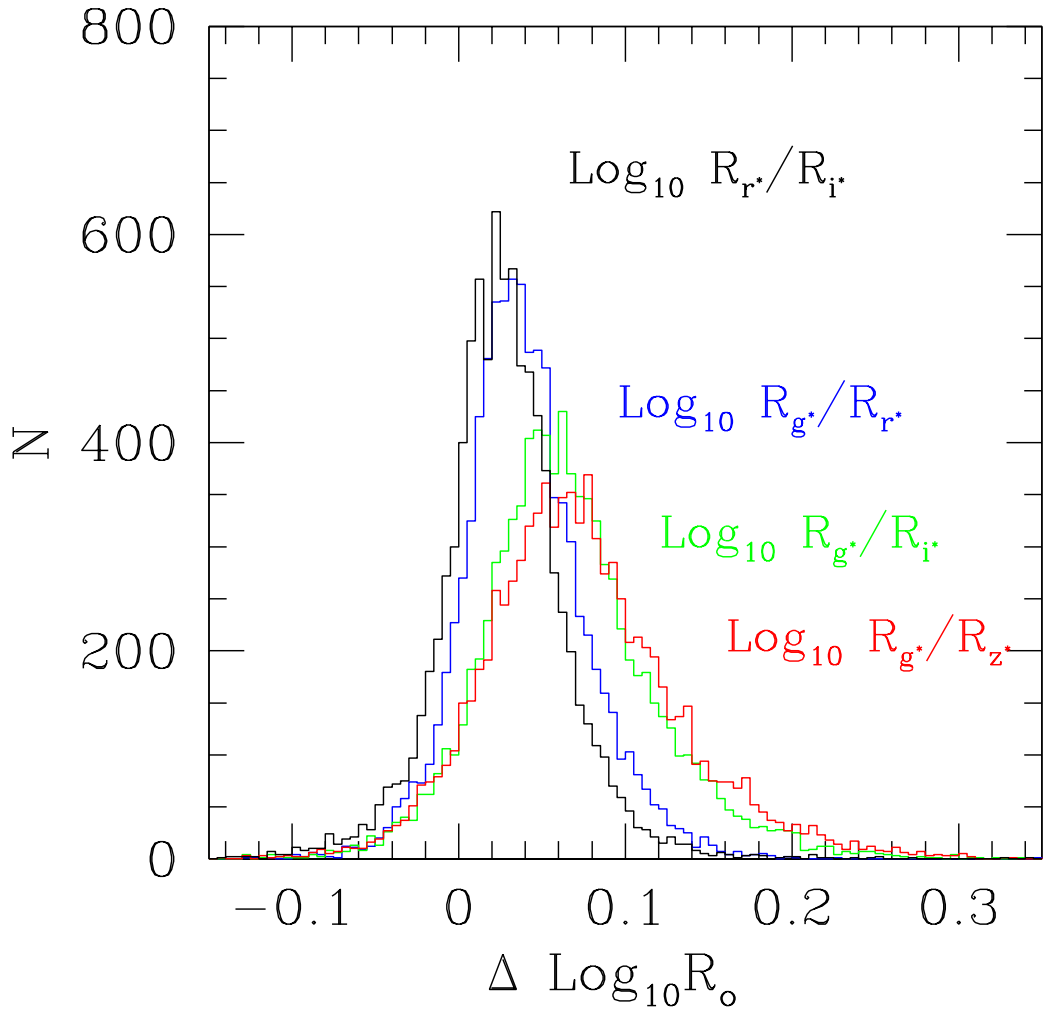}
\caption[]{Differences between the effective sizes of galaxies 
in different bands; the blue light is less centrally concentrated 
than the red light.  }
\label{rdiffs}
\end{figure}

It has been known for some time that giant early-type galaxies 
are reddest in their cores and become bluer toward their edges 
(e.g., de Vaucouleurs 1961; Sandage \& Visvanathan 1978a).  
Figure~3 of Paper~I shows that the half-light angular sizes of the 
galaxies in our sample are indeed larger in the bluer bands.  
Figure~\ref{rdiffs} shows how the effective physical radii of the 
galaxies in our sample change in the four bands.  On average, 
early-type galaxies have larger effective radii in the bluer bands.  
This trend indicates that there are color gradients in early-type 
galaxies.  The distribution of size ratios does not correlate with 
luminosity.  However, the ratio of the effective size in the 
$g^*$ and $r^*$ bands is slightly larger for bluer galaxies 
than for redder ones, suggesting that color gradients are stronger 
in the galaxies which are bluer.  In addition the scatter around 
the mean ratio is slightly larger for the bluer galaxies.  

As Scodeggio (2001) emphasizes, if the effective sizes of galaxies 
depend on waveband, then the strength of the color-magnitude relation 
depends on how the color is defined.  
Therefore, we have tried five different definitions for the color.  
The first uses the total luminosities one infers from fitting a 
de Vaucouleurs model to the light in a given band.  The `total' 
colors defined in this way are relatively noisy, because they depend 
on independent fits to the surface brightness distributions in each 
band (c.f. discussion in Paper~I).  
Since the half-light radius is larger in the bluer bands, a greater 
fraction of the light in the redder bands comes from regions which are 
closer to the center than for the bluer bands.  Therefore, this total 
color can be quite different from that which one obtains with a 
fixed angular or physical aperture.  

To approximate fixed physical aperture colors, we have integrated the 
de~Vaucouleurs profiles in the different bands assuming a tophat 
filter (since this can be done analytically) of scale $f$ times the 
effective $r^*$ radius, $R_o(r^*)$, for a few choices of $f$.  
The resulting colors depend on $f$, and the slope of the associated 
color magnitude relation decreases as $f$ decreases.  We have 
arbitrarily chosen to present results for $f=2$.  
These are not quite fixed aperture colors, since the effective angular 
aperture size varies from one galaxy to another, but, for any given 
galaxy, the aperture size is the same in all the bands (i.e., it is 
related to the effective radius in $r^*$).  

A third color is obtained by using the light within a fixed angular 
aperture which is the same for all galaxies.  The `fiber' magnitudes 
output by the SDSS photometric pipeline give the integrated light 
within a three arcsec aperture, and we use these to define the 
`fiber' color.  

A fourth color is that computed from the Petrosian magnitudes output 
by the SDSS photo pipeline (Stoughton et al. 2002).  

A fifth color uses the `model' magnitudes output by the SDSS photometric 
pipeline.  These are close to what one might call fixed aperture colors, 
because they are obtained by finding that filter which, given the 
signal-to-noise ratio, optimally detects the light in the $r^*$ band, 
and then using that same filter to measure the light in the other bands  
(which is one reason why they are less noisy than the total color defined 
above).  
(By definition, the model and total de~Vaucouleurs magnitudes are the 
same in $r^*$.  They are different in other bands because the effective 
radius is a function of wavelength.  We have verified that the difference 
between these two magnitudes in a given band correlates with the difference 
between the effective radius in $r^*$ and the band in question.)
In this respect, the model colors are similar to those one might get with 
a fixed physical aperture (they would be just like the fixed physical 
aperture colors if the optimal smoothing filter was a tophat).  These, 
also, are not fixed angular aperture colors, since the effective aperture 
size varies from one galaxy to another, but, for any given galaxy, the 
aperture is the same in all the bands.  

\begin{figure}
\centering
\epsfxsize=\hsize\epsffile{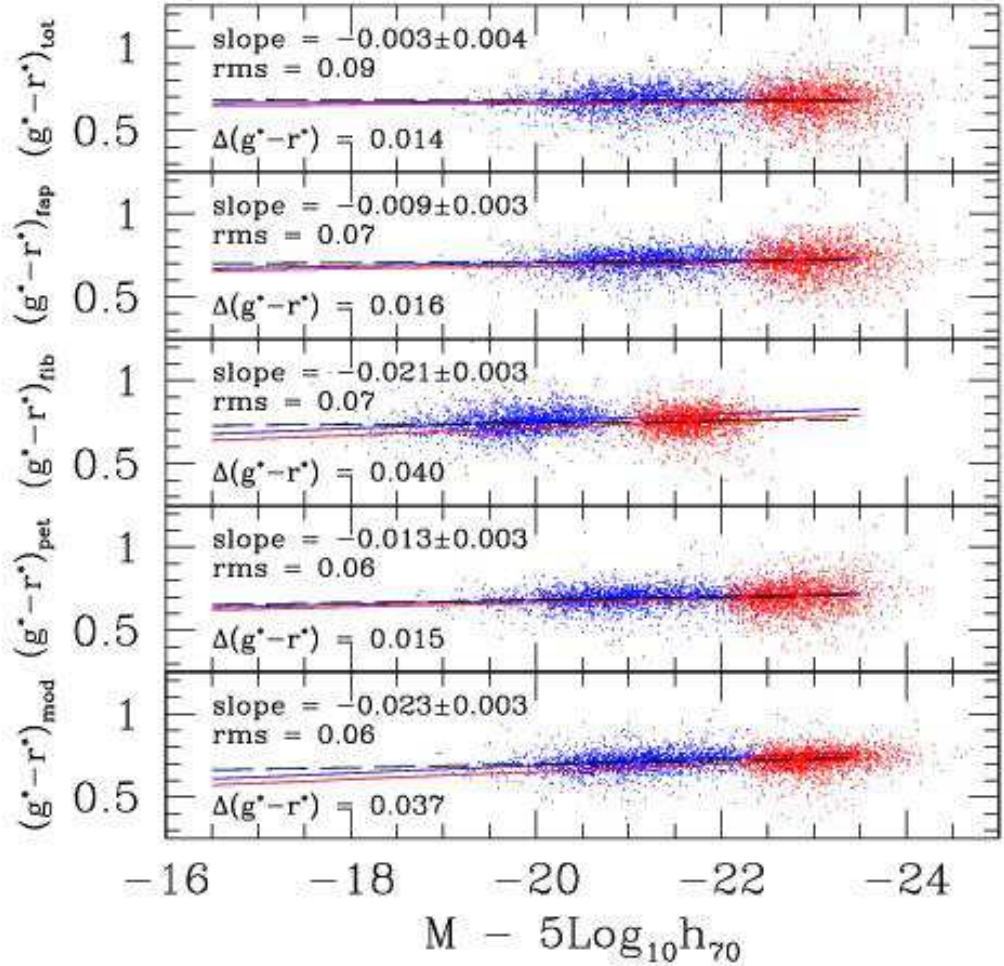}
\caption{Color-magnitude relations associated with various definitions 
of magnitude and $g^*-r^*$ color.  Top-left of each panel shows the 
slope determined from a low redshift subsample.  Fixed-aperture colors 
(bottom panel) give steeper color--magnitude relations; the correlation 
is almost completely absent if colors are defined using the total 
magnitudes (top panel).  Bottom left of each panel shows the zero-point 
shift required to fit the higher redshift sample.  This shift is an 
estimate of how the colors have evolved---it, too, depends on how the 
color was defined.  Dashed lines show fits to the whole sample; because 
they ignore the evolution of the colors, they are significantly 
shallower than fits which are restricted to a small range in redshifts, 
for which neglecting evolution is a better approximation. }  
\label{fig:cmag}
\end{figure}

\begin{figure}
\centering
\epsfxsize=\hsize\epsffile{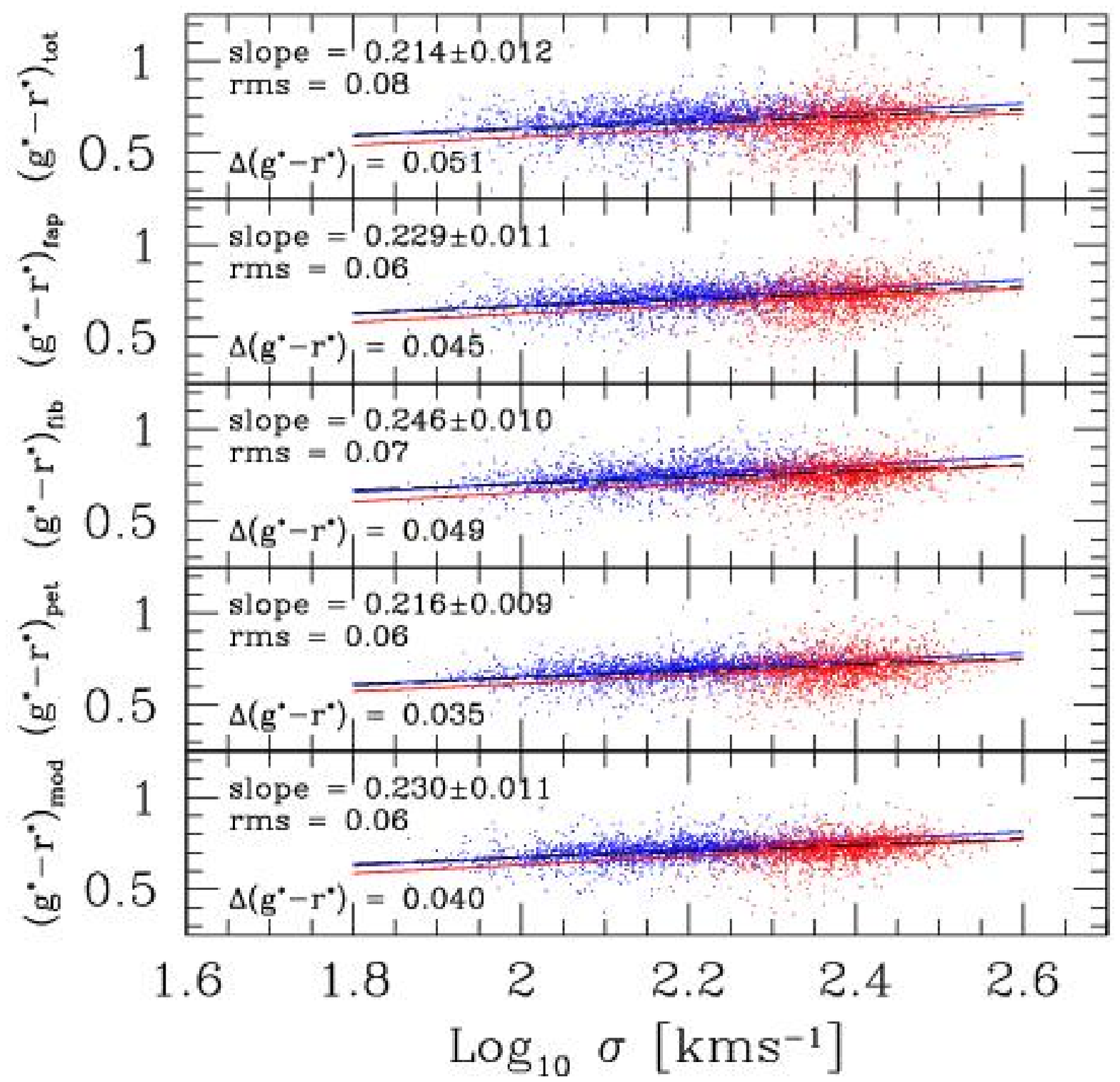}
\caption{As for the previous figure, but now showing how the 
color-$\sigma$ relation varies as the definition of $g^*-r^*$ color
changes.}
\label{fig:csig}
\end{figure}

A final possibility is to use `spectral magnitudes'; these can be made 
by integrating up the light in the spectrum of each galaxy, weighting by 
the different pass-band filters.  Whereas the other five colors require 
a good understanding of the systematics of the photometric data sets, 
this one requires a similar understanding of the spectroscopic data sets 
also.  We have not done this here, although it should be possible in 
the near future.  

The resulting $g^*-r^*$ color-magnitude relations are shown in 
Figure~\ref{fig:cmag}.  
The $x$-axis in the top two panels is the de Vaucouleurs magnitude in 
$r^*$, whereas it is the fiber magnitude in $r^*$ in the third panel, 
the Petrosian $r^*$ magnitude in the fourth panel, 
and the model $r^*$ magnitude in the bottom panel.  
So that evolution effects do not combine with the magnitude limit 
of our sample to produce a shallow relation, we divided our sample into 
two: a low-redshift sample, which includes all galaxies at $z\le 0.08$, 
and a high-redshift $z\ge 0.16$ sample.  For each definition of color, 
we computed the slope and amplitude of the color--magnitude relation 
in the low redshift sample.  This slope is shown in the top left corner 
of each panel.  We then required the slope of the high redshift sample 
to be the same (recall from Figure~\ref{cmag3} that this is a good 
approximation); the offset required to get a good fit is shown in 
the bottom left of each panel.  This is the quantity which provides 
an estimate of how much the colors have evolved.  The two thin solid 
lines in each panel show the low- and high-redshift color--magnitude 
relations computed in this way.  For comparison, the dashed line shows 
a fit to the full sample, ignoring evolution effects; in all the 
panels, it is obviously much flatter than the relation at low redshift.  

The figure shows clearly that the slope of the color-magnitude 
relation depends on how the color was defined:  
it is present when fixed-apertures are used (e.g., bottom panel), 
and it is almost completely absent when the total light within the 
de Vaucouleurs fit is used (top panel).  Our results are consistent 
with those reported by Okamura et al. (1998) and Scodeggio (2001).  
Note that one's inference of how much the colors have evolved,  
$\Delta (g^*-r^*)$, also depends on how the color was defined.  

A similar comparison for the correlation between color and velocity 
dispersion $\sigma$ is presented in Figure~\ref{fig:csig}.  
We have already argued that color$-\sigma$ is the primary correlation; 
this relation is also considerably less sensitive to the different 
definitions of color.  However, it is sensitive to evolution:  a fit 
to the full sample gives a slope of 0.14, compared to the value of 0.23 
for the low redshift sample.  Because the mean color$-\sigma$ relation 
is steeper than that between color and magnitude, the change to the 
slope of the relation is less dramatic.  The zero-point shifts, which 
estimate the evolution of the color, are comparable both for the 
color--magnitude and the color$-\sigma$ relations, provided the SDSS 
model colors are used (bottom panel).

\section{Line-indices: Chemical evolution and environment}\label{lindices}
We now turn to a more detailed study of the spectra in our sample.  
To measure spectral features reliably requires a spectrum with a higher 
signal-to-noise ratio than we have for any individual galaxy in our 
sample.  Section~\ref{coadd} describes the procedure we have adopted 
to deal with this.  All the line indices we study below correlate with 
velocity dispersion $\sigma$.  Because $\sigma$ correlates with 
luminosity, the magnitude limit of our sample means that we have no 
objects with low velocity dispersions at high redshifts.  
By presenting results at fixed velocity dispersion, our analysis of 
line indices should not be biased by this selection:  this is the 
subject of Section~\ref{fixedv}.

\subsection{Composite spectra of similar objects}\label{coadd}

The typical signal-to-noise ratio of the spectra in our sample is 
about 15 (Figure~18 in Paper~I).  To measure spectral features reliably 
requires a spectrum with $S/N\sim 100$ (e.g. Trager et al. 1998), 
so we have adopted the following procedure.  

\begin{figure}
\centering
\epsfxsize=\hsize\epsffile{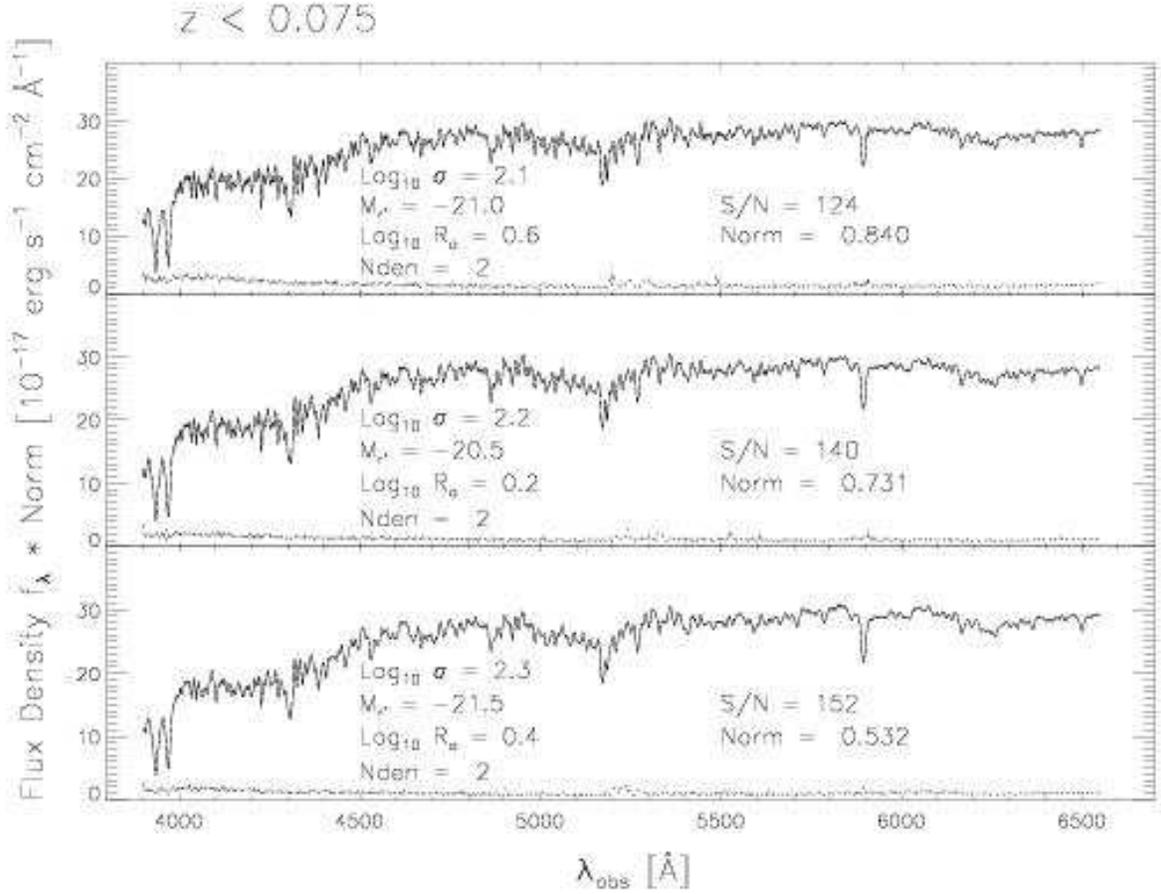}
\caption{Composite spectrum obtained by co-adding the spectra of 
galaxies with similar redshifts, velocity dispersions, 
absolute magnitudes, effective radii and local densities.  
The line at the bottom of each panel shows the rms scatter of the 
individual spectra used to make the composite spectrum.
The signal-to-noise ratio of the composite spectrum is also shown.  }
\label{fig:composites}
\end{figure}

We have a large number of galaxies in our sample, many of which have 
similar luminosities, sizes, velocity dispersions, and redshifts.  
By co-adding the spectra of similar galaxies, we can produce a 
composite spectrum with a considerably higher $S/N$ ratio.  Since we 
wish to increase the $S/N$ ratio by a factor of about seven, 
we need at least fifty galaxies per composite spectrum.  On the other 
hand, we do not want to co-add spectra of galaxies which differ 
considerably from each other.  Therefore, we divided the galaxies 
in our sample into five bins each of redshift, luminosity, velocity 
dispersion, effective radius, and density, and co-added the spectra of 
all the galaxies in each bin.  This gave about 200 composite spectra, 
with varying numbers of galaxies contributing to each.  We then 
excluded from further consideration all composites which had $S/N<50$.
Figure~\ref{fig:composites} shows a selection of the 182 composite 
spectra with $S/N>50$ (for given bins in $\sigma$ and $z$, we show 
the composite spectrum which has the median $S/N$ ratio). 
The line at the bottom of each panel shows the rms scatter of the 
individual spectra used to make the composite spectrum.  
These 182 composite spectra, the scatter, and the errors, as a
function of restframe wavelength are available electronically; 
interested readers should contact the first author directly.  

\begin{figure}
\centering
\epsfxsize=\hsize\epsffile{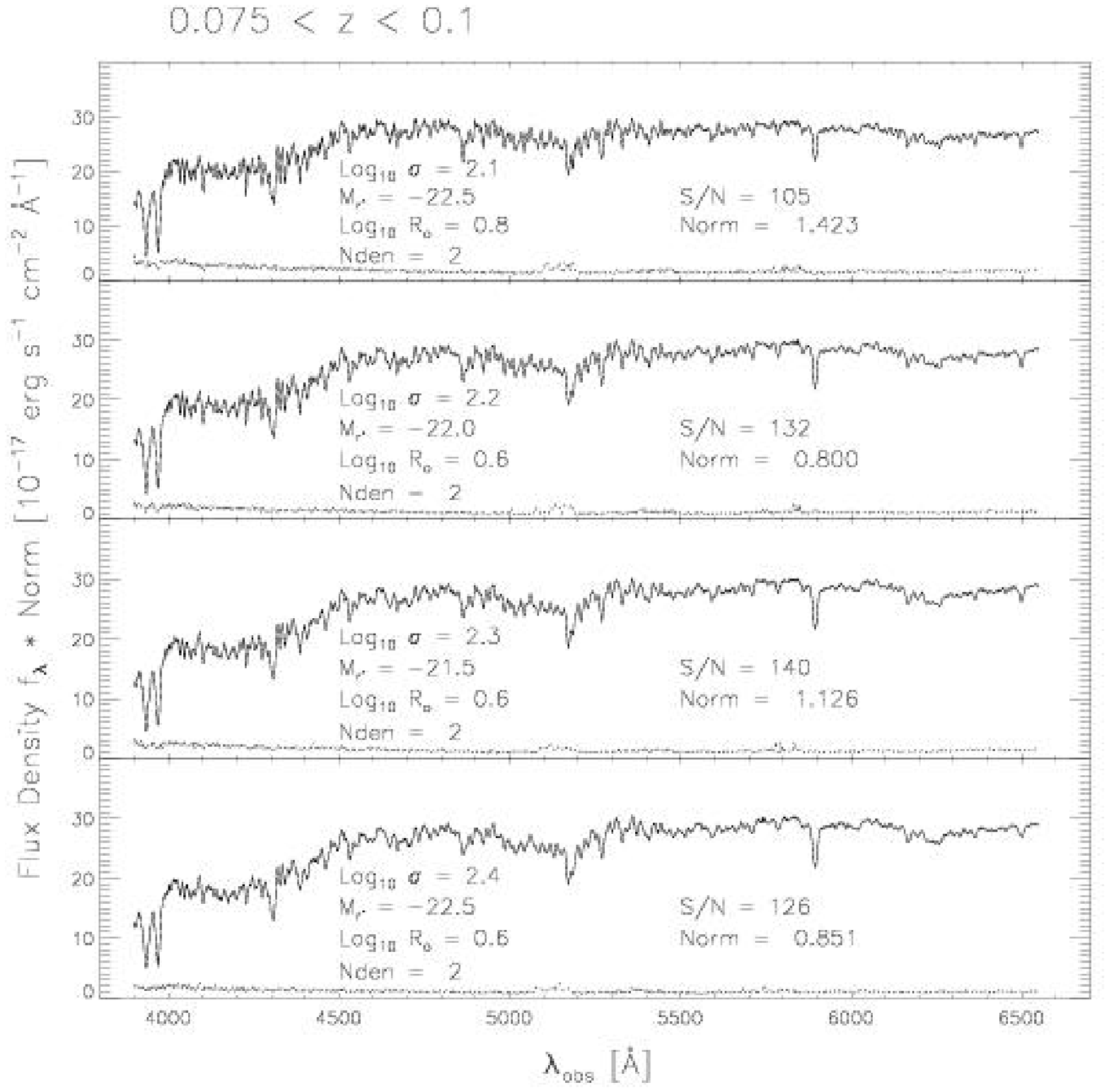}
\begin{center}
Fig.~\ref{fig:composites}. -- Continued.
\end{center}
\end{figure}

\begin{figure}
\centering
\epsfxsize=\hsize\epsffile{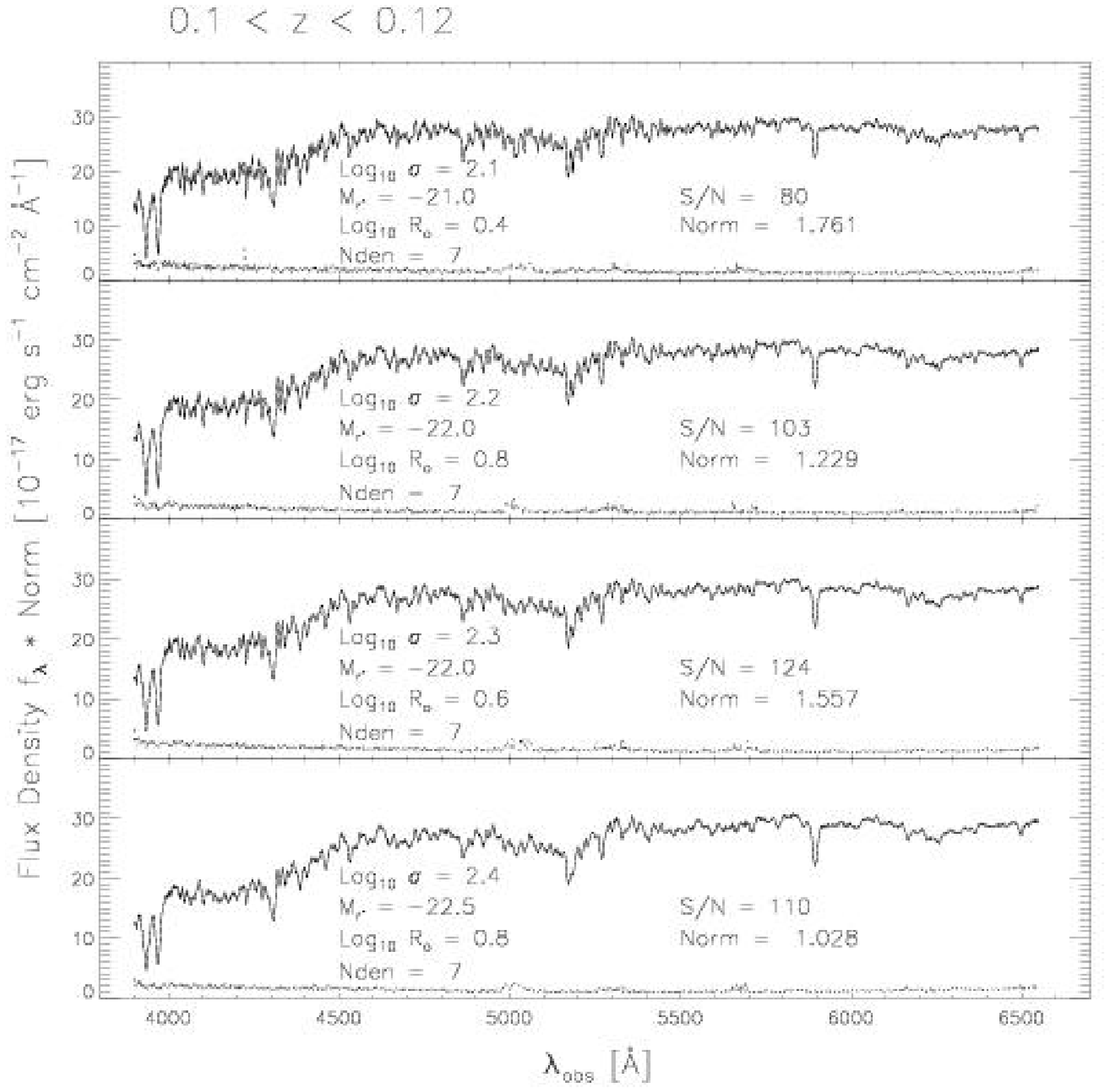}
\begin{center}
Fig.~\ref{fig:composites}. -- Continued.
\end{center}
\end{figure}

\begin{figure}
\centering
\epsfxsize=\hsize\epsffile{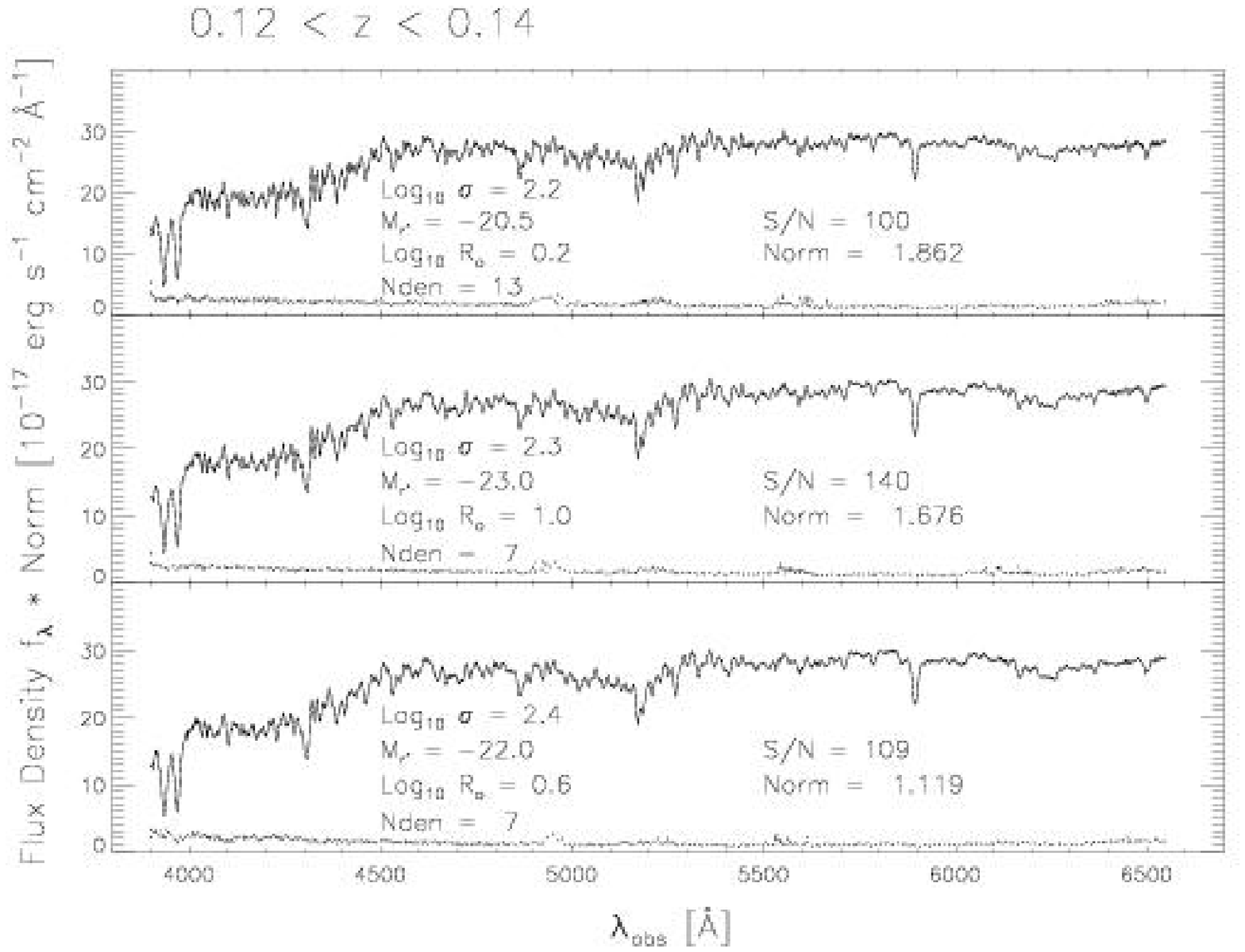}
\begin{center}
Fig.~\ref{fig:composites}. -- Continued.
\end{center}
\end{figure}

\begin{figure}
\centering
\epsfxsize=\hsize\epsffile{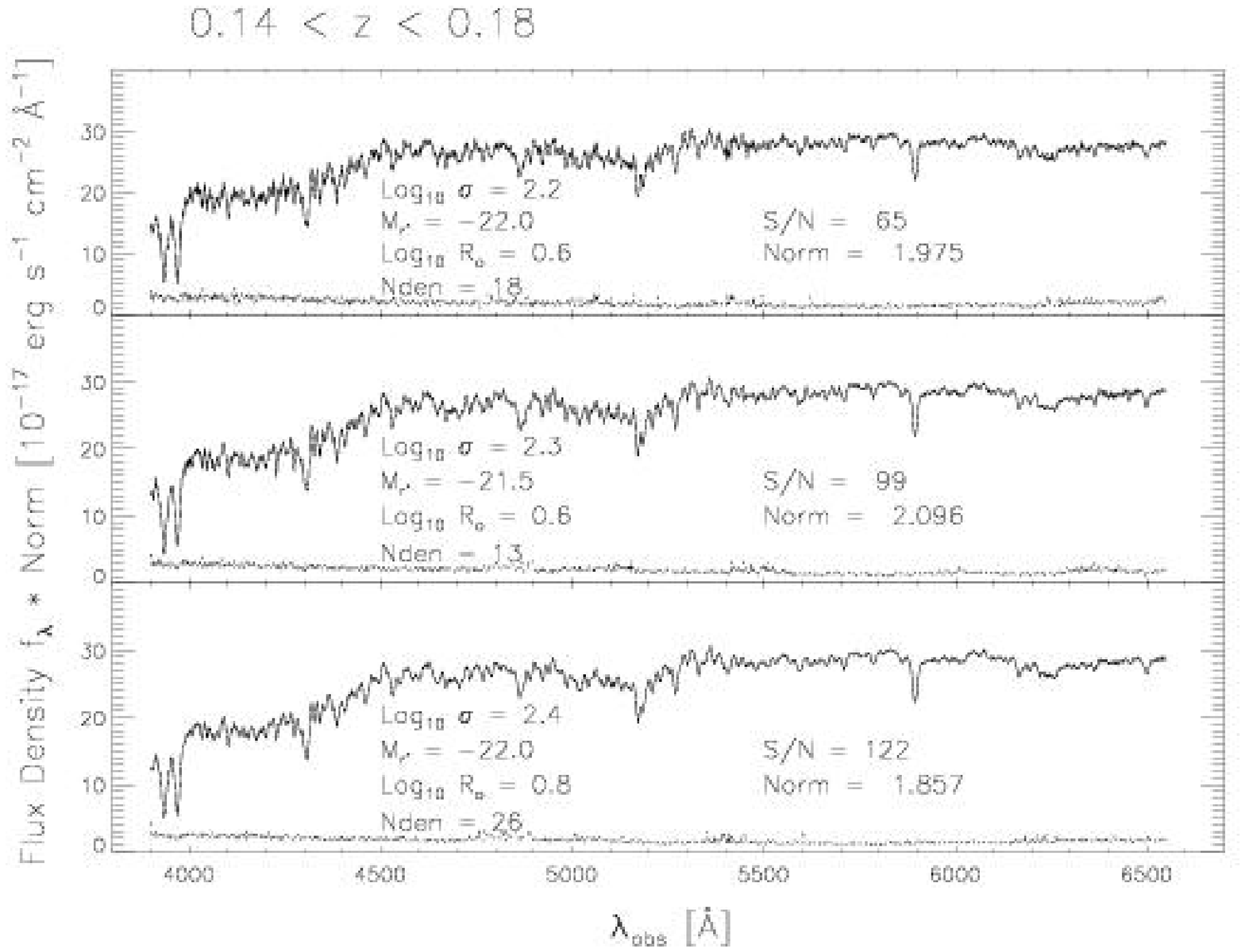}
\begin{center}
Fig.~\ref{fig:composites}. -- Continued.
\end{center}
\end{figure}

\begin{figure}
\centering
\epsfxsize=\hsize\epsffile{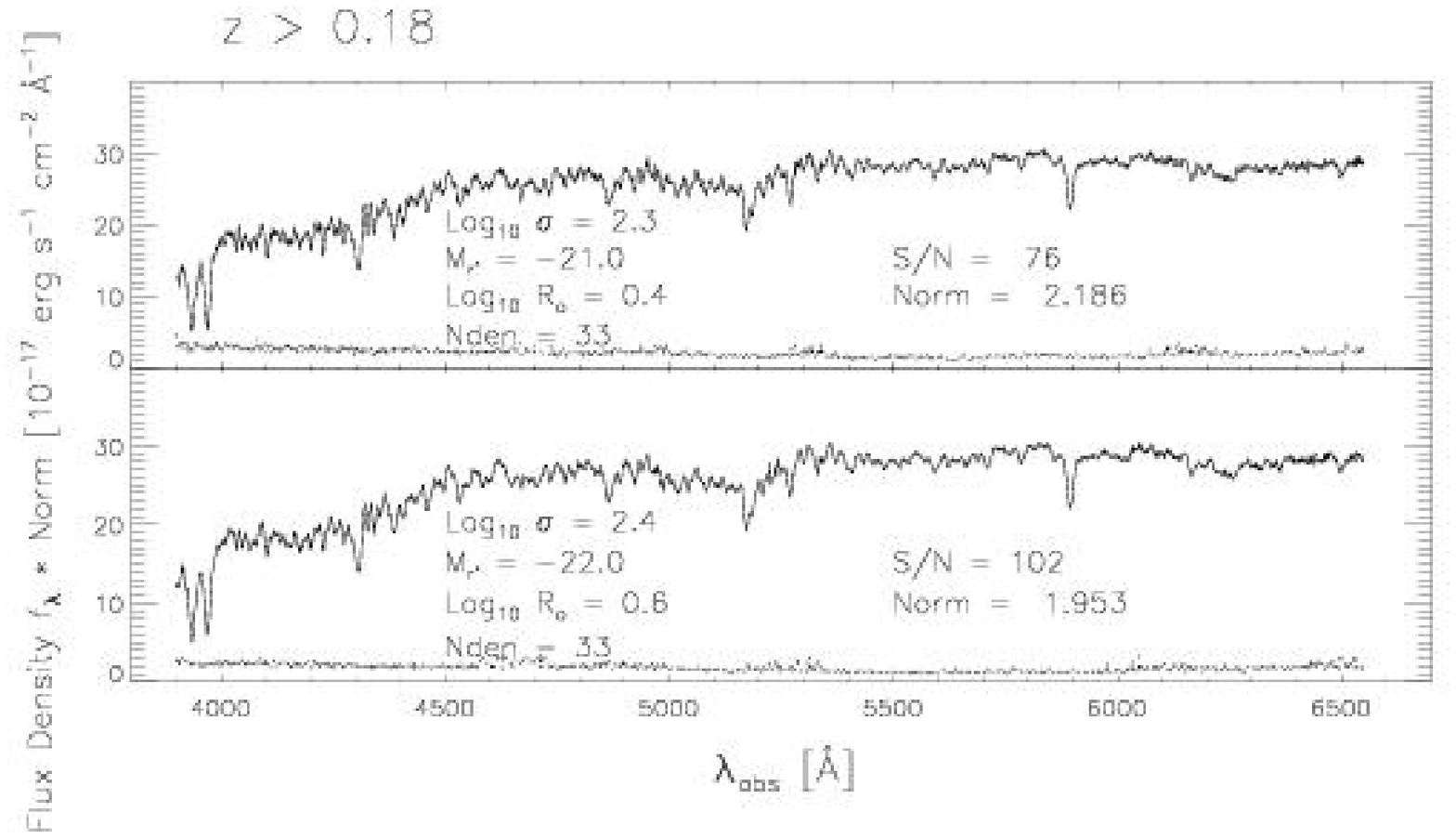}
\begin{center}
Fig.~\ref{fig:composites}. -- Continued.
\end{center}
\end{figure}

We estimated the Mg$_2$, Mg$b$, H$_\beta$ and $\langle{\rm Fe}\rangle$ 
line-indices in the higher signal-to-noise composite spectra following 
methods outlined by Trager et al. (1998).  (Analysis of the properties 
of early-type galaxies using these higher signal-to-noise composite 
spectra is on-going.)  The estimated indices were aperture-corrected 
following J{\o}rgensen (1997):
Mg$_2$ = Mg$_2^{\rm est}$ + $0.04\log_{10}[1.5/(r_o/8)]$, 
${\rm H}_\beta = {\rm H}_\beta^{\rm est} [1.5/(r_o/8)]^{-0.005}$, and 
$\langle{\rm Fe}\rangle = 
    \langle{\rm Fe}\rangle^{\rm est}[1.5/(r_o/8)]^{0.05}$ and 
$\langle{\rm Mg}b\rangle = 
    \langle{\rm Mg}b\rangle^{\rm est}[1.5/(r_o/8)]^{0.05}$.  
(Because the indices were measured for co-added spectra, we use the mean 
values of $r_o$ in each bin to make the aperture correction.)  
In addition, the observed line indices of an individual galaxy are 
broadened by the velocity dispersion of the galaxy.  Simulations similar 
to those we used to estimate the velocity dispersion itself 
(see Appendix~B in Paper~I) were used to estimate and correct for 
the effect of the broadening.  
For all the indices presented here, the required corrections increase 
with increasing $\sigma$.  (We use the mean value of $\sigma$ in 
each bin to make the corrections.)
Whereas the corrections to Mg$_2$ and H$_\beta$ are small (on the order 
of a percent), the corrections to Mg$b$ and Fe are larger (on the order 
of ten percent).  

Table~\ref{tab:indices} summarizes our line-index measurements.  
Column 1) gives the ID number of the composite spectrum; 
columns 2-6) give the centers of the bins in velocity dispersion, 
size, absolute magnitude, redshift and local density which were 
used to define which galaxies contribute to the composite; and 
columns 7-14) give the measured strength of the index and the 
associated error on the measurement, for H$_\beta$, Mg$_2$, Mg$_b$ 
and $\langle{\rm Fe}\rangle$.  
 
\renewcommand{\arraystretch}{.6}
\begin{deluxetable}{cccccccccccccc}
\rotate
\tablewidth{0pc}
\tablecaption{Line--index measurements from co-added spectra.}
\tablehead{
\colhead{ID} & \colhead{$\log \sigma$} & \colhead{$\log R_o$} 
     & \colhead{$M_{r^*}$}
     & \colhead{$z$} & \colhead{Nden} 
     & \colhead{$\log$H$_{\beta}$} & \colhead{$\delta\log$H$_{\beta}$}   
     & \colhead{Mg$_2$} & \colhead{$\delta$Mg$_2$}  
     & \colhead{$\log$Mg$b$} & \colhead{$\delta\log$Mg$b$}
     & \colhead{$\log \langle{\rm Fe}\rangle$}  
     & \colhead{$\delta \log \langle{\rm Fe}\rangle$}
\\
\colhead{}  & \colhead{[km~s$^{-1}$]} & \colhead{[$h_{70}^{-1}$~kpc]} & \colhead{mag} & \colhead{} & \colhead{} 
     & \colhead{[\AA]} & \colhead{[\AA]} & \colhead{mag} & \colhead{mag} 
     & \colhead{[\AA]} & \colhead{[\AA]} & \colhead{[\AA]} & \colhead{[\AA]}
}
\startdata
  1 &  2.10 &  0.20 &  -20.50 &  0.08 &   2 &  0.274 &  0.004 &  0.259 &  0.001 &  0.607 &  0.002 &  0.481 &  0.005\\
  2 &  2.10 &  0.40 &  -20.50 &  0.08 &   2 &  0.219 &  0.005 &  0.264 &  0.001 &  0.594 &  0.002 &  0.490 &  0.002\\
  3 &  2.10 &  0.60 &  -21.00 &  0.08 &   2 &  0.256 &  0.005 &  0.251 &  0.002 &  0.609 &  0.003 &  0.515 &  0.005\\
  4 &  2.10 &  0.60 &  -21.50 &  0.08 &   2 &  0.242 &  0.005 &  0.260 &  0.002 &  0.586 &  0.003 &  0.489 &  0.002\\
  5 &  2.20 &  0.20 &  -20.50 &  0.08 &   2 &  0.209 &  0.005 &  0.301 &  0.001 &  0.664 &  0.002 &  0.456 &  0.002\\
  6 &  2.20 &  0.40 &  -20.50 &  0.08 &   2 &  0.212 &  0.007 &  0.284 &  0.002 &  0.661 &  0.003 &  0.471 &  0.002\\
  7 &  2.20 &  0.40 &  -21.00 &  0.08 &   2 &  0.212 &  0.003 &  0.286 &  0.001 &  0.660 &  0.001 &  0.496 &  0.003\\
  8 &  2.20 &  0.40 &  -21.50 &  0.08 &   2 &  0.244 &  0.004 &  0.281 &  0.001 &  0.637 &  0.002 &  0.502 &  0.006\\
  9 &  2.20 &  0.60 &  -21.00 &  0.08 &   2 &  0.167 &  0.007 &  0.267 &  0.001 &  0.622 &  0.002 &  0.451 &  0.005\\
 10 &  2.20 &  0.60 &  -21.50 &  0.08 &   2 &  0.199 &  0.004 &  0.273 &  0.001 &  0.652 &  0.002 &  0.519 &  0.006\\
 11 &  2.20 &  0.60 &  -22.00 &  0.08 &   2 &  0.261 &  0.006 &  0.269 &  0.001 &  0.600 &  0.003 &  0.491 &  0.003\\
 12 &  2.20 &  0.80 &  -22.00 &  0.08 &   2 &  0.151 &  0.007 &  0.250 &  0.001 &  0.606 &  0.002 &  0.474 &  0.003\\
 13 &  2.30 &  0.40 &  -21.00 &  0.08 &   2 &  0.158 &  0.006 &  0.314 &  0.002 &  0.686 &  0.002 &  0.521 &  0.007\\
 14 &  2.30 &  0.40 &  -21.50 &  0.08 &   2 &  0.182 &  0.004 &  0.312 &  0.001 &  0.673 &  0.002 &  0.485 &  0.003\\
 15 &  2.30 &  0.60 &  -21.50 &  0.08 &   2 &  0.115 &  0.010 &  0.297 &  0.003 &  0.691 &  0.004 &  0.476 &  0.007\\
 16 &  2.30 &  0.60 &  -22.00 &  0.08 &   2 &  0.174 &  0.004 &  0.291 &  0.001 &  0.665 &  0.002 &  0.483 &  0.004\\
 17 &  2.10 &  0.40 &  -21.00 &  0.10 &   2 &  0.263 &  0.005 &  0.249 &  0.003 &  0.615 &  0.005 &  0.475 &  0.005\\
 18 &  2.10 &  0.60 &  -21.00 &  0.10 &   2 &  0.234 &  0.007 &  0.253 &  0.003 &  0.617 &  0.006 &  0.496 &  0.005\\
 19 &  2.10 &  0.60 &  -21.50 &  0.10 &   2 &  0.253 &  0.005 &  0.235 &  0.002 &  0.581 &  0.005 &  0.480 &  0.002\\
 20 &  2.20 &  0.40 &  -21.00 &  0.10 &   2 &  0.218 &  0.004 &  0.281 &  0.002 &  0.676 &  0.003 &  0.497 &  0.004\\
 21 &  2.20 &  0.40 &  -21.50 &  0.10 &   2 &  0.274 &  0.005 &  0.246 &  0.002 &  0.606 &  0.004 &  0.480 &  0.002\\
 22 &  2.20 &  0.60 &  -21.50 &  0.10 &   2 &  0.226 &  0.004 &  0.246 &  0.002 &  0.606 &  0.003 &  0.492 &  0.003\\
 23 &  2.20 &  0.80 &  -22.00 &  0.10 &   2 &  0.215 &  0.005 &  0.255 &  0.002 &  0.634 &  0.003 &  0.482 &  0.002\\
 24 &  2.30 &  0.40 &  -21.00 &  0.10 &   2 &  0.123 &  0.007 &  0.304 &  0.002 &  0.734 &  0.003 &  0.515 &  0.003\\
 25 &  2.30 &  0.60 &  -21.50 &  0.10 &   2 &  0.167 &  0.005 &  0.296 &  0.002 &  0.692 &  0.002 &  0.482 &  0.004\\
 26 &  2.30 &  0.60 &  -22.00 &  0.10 &   2 &  0.213 &  0.003 &  0.277 &  0.001 &  0.672 &  0.002 &  0.498 &  0.003\\
 27 &  2.30 &  0.80 &  -22.00 &  0.10 &   2 &  0.172 &  0.005 &  0.252 &  0.003 &  0.606 &  0.005 &  0.472 &  0.005\\
 28 &  2.30 &  0.80 &  -22.50 &  0.10 &   2 &  0.216 &  0.004 &  0.271 &  0.002 &  0.692 &  0.002 &  0.487 &  0.003\\
 29 &  2.10 &  0.60 &  -21.50 &  0.12 &   2 &  0.273 &  0.008 &  0.255 &  0.003 &  0.580 &  0.005 &  0.485 &  0.006\\
 30 &  2.20 &  0.60 &  -21.50 &  0.12 &   2 &  0.244 &  0.004 &  0.276 &  0.001 &  0.643 &  0.002 &  0.491 &  0.005\\
\enddata
\tablecomments{The complete version of this table is in the electronic
edition of the Journal.  The printed edition contains only a sample.}
\label{tab:indices}
\end{deluxetable}

\subsection{Correlations with velocity dispersion}\label{fixedv}

\begin{figure}[t]
\centering
\epsfxsize=\hsize\epsffile{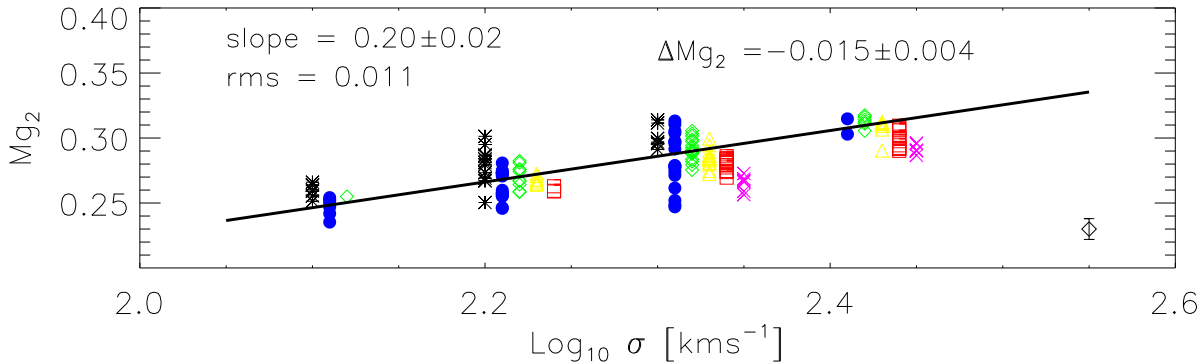}
\vspace{-10cm}
\caption{Mg$_2$ as a function of $\sigma$.  
Stars, filled circles, diamonds, triangles, squares and crosses show 
results from coadded spectra of similar galaxies in successively higher 
redshift bins ($z<0.075$, $0.075<z\le 0.1$, $0.1<z\le 0.12$, 
$0.12<z\le 0.14$, $0.14<z\le 0.18$, and $z>0.18$).  
Symbol with bar in bottom corner shows the typical uncertainty on the 
measurements.  At fixed redshift, Mg$_2$ increases with increasing 
$\sigma$.  At fixed $\sigma$, the spectra from higher redshift galaxies 
are weaker in Mg$_2$.  Text at top right shows the shift between the 
lowest and highest redshift bins averaged over the mean shifts at 
$\log_{10}\sigma=2.2$, 2.3 and 2.4.  
We also performed linear fits to the relations at each redshift, and then 
averaged the slopes, zero-points and rms scatter around the fit.  
Solid line shows the mean relation obtained in this way, and text at 
top shows the averaged slope and averaged scatter.  }
\label{fig:Mg2sigma}
\end{figure}

\begin{figure}
\centering
\epsfxsize=\hsize\epsffile{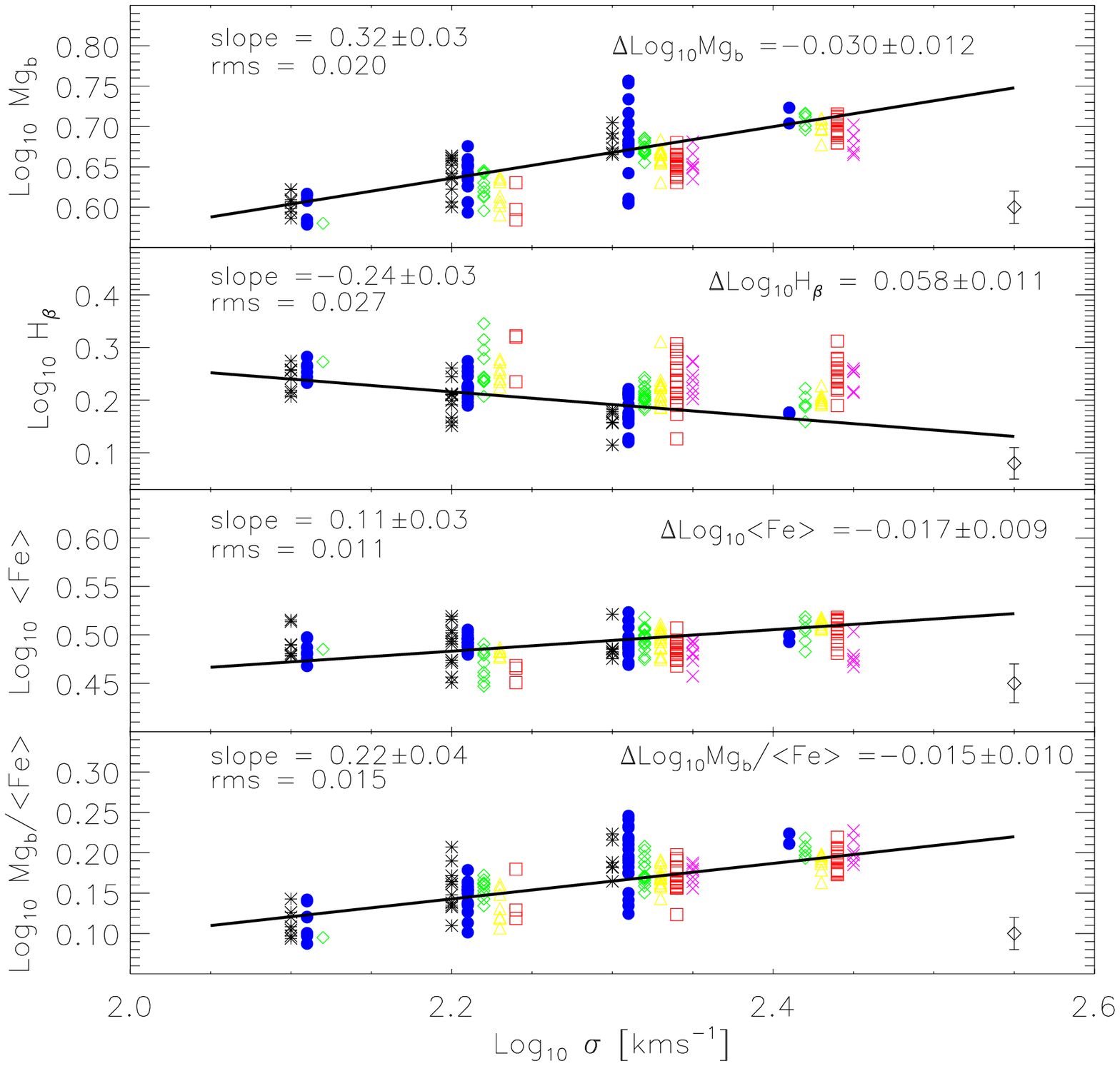}
\caption{Same as previous figure, but now showing the spectral line-indices 
Mg$b$, H$_\beta$, $\langle{\rm Fe}\rangle$, and the ratio [Mg$b$/Fe] 
(top to bottom) as functions of $\sigma$.  
At fixed redshift, Mg$b$ and $\langle{\rm Fe}\rangle$ increase, 
whereas H$_\beta$ decreases with increasing $\sigma$.  
At fixed $\sigma$, the spectra from higher redshift galaxies are weaker 
in both Mg$_2$ and $\langle{\rm Fe}\rangle$, but stronger in H$_\beta$.  
Text at top right shows the shift between the lowest and highest redshfit 
bins averaged over the values at $\log_{10}\sigma=2.2$, 2.3 and 2.4.  }
\label{fig:lindices}
\end{figure}

Figures~\ref{fig:Mg2sigma} and~\ref{fig:lindices} show how the 
line-indices in Table~\ref{tab:indices} correlate with velocity dispersion.  
In all panels, stars, filled circles, diamonds, triangles, squares and 
crosses show the redshift bins $z<0.075$, $0.075<z\le 0.1$, $0.1<z\le 0.12$, 
$0.12<z\le 0.14$, $0.14<z\le 0.18$, and $z>0.18$.  The median redshifts 
in these bins are 0.062, 0.086, 0.110, 0.130, 0.156 and 0.200.  For 
clarity, at each bin in velocity dispersion, the symbols for successive 
redshift bins have been offset slightly to the right from each other.  
This should help to separate out the effects of evolution from those which 
are due to the correlation with $\sigma$.  The solid line and text in each 
panel shows the relation which is obtained by performing simple linear 
fits at each redshift, and then averaging the slopes, zero-points, 
and rms scatter around the fit at each redshift.  Text at top right of 
each panel shows the shift between the lowest and highest redshift 
bins, averaged over the values at $\log_{10}\sigma=2.2$, 
$\log_{10}\sigma=2.3$ and $\log_{10}\sigma=2.4$.  Roughly speaking, 
this means that the shifts occur over a range of about $0.2-0.06=0.14$ 
in redshift, which corresponds to a time interval of 1.63~Gyr.  

Figure~\ref{fig:Mg2sigma} shows the well-known correlation between 
Mg$_2$ and $\sigma$: at fixed redshift, 
${\rm Mg}_2\propto \sigma^{0.20\pm 0.02}$ with a scatter around the 
mean relation at each redshift of 0.011~mags.  The fit we find is similar 
to that found in previous work based on spectra of individual (as opposed 
to coadded) galaxies (e.g., J{\o}rgensen 1997; Bernardi et al. 1998; 
Pahre et al. 1998; Kuntschner 2000; Blakeslee et al. 2001; 
Bernardi et al. 2002), although the scatter we find is somewhat 
smaller.  The slope of our fit is shallower than that reported by 
Colless et al. (1999), but this may be consequence of our decision to 
perform linear regression, rather than maximum-likelihood, fits.  
(Maximum-likelihood fits are difficult at the present time because 
our bins in luminosity are rather large.  We plan to make the 
maximum-likelihood estimate when the sample is larger, so that finer 
bins in luminosity can be made.)  

Although the magnitude limit of our sample makes it difficult to study 
the evolution of the Mg$_2 - \sigma$ relation, a few bins in $\sigma$ 
do have galaxies from a range of different redshifts.  Recall that, for 
the purposes of presentation, the points in each bin in $\sigma$ have been 
shifted to the right by an amount which depends on the redshift bin they 
represent.  When plotted in this way, the fact that the points associated 
with each bin in $\sigma$ slope down and to the right suggests that, at 
fixed $\sigma$, the higher redshift galaxies have smaller values of Mg$_2$.  
Large values of Mg$_2$ are expected to indicate either that the stellar 
population is metal rich, or old, or both.  Thus, in a passively evolving 
population, the relation should be weaker at high redshift.  
This is consistent with the trend we see.  The average value of Mg$_2$ 
decreases by about $(0.015\pm 0.004)$~mags between our lowest and highest 
redshift bins (a range of about 1.63$h^{-1}$Gyr).  
We will return to this shortly.  

The top panel of Figure~\ref{fig:lindices} shows that, at fixed redshift, 
Mg$b\propto \sigma^{0.32\pm 0.03}$, with a scatter of 0.020.  This is 
consistent with the scaling reported by Trager et al. (1998).  
[A plot of Mg$b$ versus Mg$_2$ is well fit by 
$\log_{10}{\rm Mg}b = (1.41\pm 0.18){\rm Mg}_2 + 0.26$; 
this slope is close to the value $0.32/0.20$ one estimates from the 
individual Mg$b-\sigma$ and Mg$_2-\sigma$ relations.  It is also 
consistent with Figure~58 in Worthey (1994).]  
As was the case for Mg$_2$, our data indicate that, at fixed velocity 
dispersion, Mg$b$ is weaker in the higher redshift population.  
The average difference between our lowest and highest redshift bins is 
$0.030\pm 0.012$.  This corresponds to a fractional change in Mg$b$ of 
0.07 over about 1.63$h^{-1}$Gyr.  In contrast 
Bender, Ziegler \& Bruzual (1996) find that Mg$b$ at $z=0.37$ is smaller 
by 0.3\AA\ compared to the value at $z=0$.  This is a fractional change 
of about 0.07 but over a redshift range which corresponds to a time 
interval of 4$h^{-1}$Gyr.  Bender et al. also reported weak evidence of 
differential evolution:  the low $\sigma$ population appeared to have 
evolved more rapidly.  Our Mg$_2-\sigma$ and Mg$b-\sigma$ relations also 
show some evidence of such a trend.  

Colless et al. (1999) define 
Mg$b' = -2.5\log_{10}(1 - {\rm Mg}b/32.5)$, and show that their 
data are well fit by 
Mg$b'\propto (0.131\pm0.017)\log_{10}\sigma - (0.131\pm 0.041)$ 
with a scatter around the mean relation of 0.022~mags.  
Kuntschner (2000) shows that the galaxies in the Fornax cluster follow 
this same scaling, although the scatter he finds is 0.011~mags.  
Our coadded spectra are also consistent with this:  we find 
Mg$b'\propto (0.15\pm 0.02)\log_{10}\sigma$, with a scatter of 0.010~mags.  
[A linear regression of the values of Mg$_2$ and Mg$b'$ in our coadded 
spectra is well fit by ${\rm Mg}_2 = (1.70\pm 0.30)\,{\rm Mg}b' - 0.01$; 
this is slightly shallower than the relation found by Colless et al.: 
${\rm Mg}_2 = 1.94{\rm Mg}b' - 0.05$.]  
The Mg$b'-\sigma$ relation in our data evolves:  in the highest 
redshift bins it is about $(0.013\pm 0.002)$~mags lower than in the lowest 
redshift bins.  Colless et al. find that in the single stellar population 
models of both Worthey (1994) and Vazdekis et al. (1996), changes in age 
or metallicity affect Mg$_2$ about twice as strongly as they do Mg$b'$.  
Figure~\ref{fig:Mg2sigma} suggests that Mg$_2$ has weakened by $-0.015$, 
so we expect Mg$b'$ to have decreased by about $-0.007$.  Therefore, 
this also suggests that the Mg$b$ (or Mg$b'$) evolution we see is large.  

The second panel in Figure~\ref{fig:lindices} shows that, at fixed 
redshift, ${\rm H}_\beta\propto \sigma^{-0.24\pm 0.03}$ with a scatter 
of 0.027.  This is consistent with J{\o}rgensen (1997), who found 
$\log_{10}{\rm H}_\beta = (-0.231\pm 0.082)\log_{10}\sigma +0.825$,
although our scatter is smaller then her value of 0.061. 
At fixed $\sigma$, H$_\beta$ is stronger 
in the higher redshift spectra.  On average, the value of H$_\beta$ 
increases by about $0.058\pm 0.011$ between our lowest and highest 
redshift bins.  
An increase of star formation activity with redshift is consistent 
with a passively evolving population.  When a larger sample is available, 
it will be interesting to see if the scatter in H$_\beta$ at fixed 
$\sigma$ also increases with redshift.  

The third panel of Figure~\ref{fig:lindices} shows that, at fixed 
redshift, $\langle{\rm Fe}\rangle\propto \sigma^{0.11\pm 0.03}$ with 
a scatter of 0.011.  This lies between the $0.075\pm 0.025$ scaling 
and scatter of 0.041 found by J{\o}rgensen (1997), and that found 
by Kuntschner (2000):  
$\langle{\rm Fe}\rangle\propto \sigma^{0.209\pm 0.047}$.  
At fixed $\sigma$, $\langle{\rm Fe}\rangle$ is slightly smaller 
at higher redshift: the change in $\log_{10}\langle{\rm Fe}\rangle$ is 
$0.017\pm 0.009$.  

The ratio Mg$b$/$\langle{\rm Fe}\rangle$ is sometimes used to 
constrain models of how early-type galaxies formed 
(e.g., Worthey, Faber \& Gonzalez 1992; Thomas, Greggio \& Bender 1999; 
but see Matteucci, Ponzone \& Gibson 1998).  In our coadded spectra, 
$\log_{10} {\rm Mg}b/\langle{\rm Fe}\rangle = 
(0.22\pm 0.04)\log_{10}\sigma - 0.34$ with a scatter of 0.015 
(bottom panel of Figure~\ref{fig:lindices}).  
The slope of this relation equals the difference between the slopes 
of the Mg$b-\sigma$ and $\langle {\rm Fe}\rangle - \sigma$ relations, 
and there is marginal evidence of evolution:  the change in 
Mg$b$/$\langle{\rm Fe}\rangle$ is $0.015\pm 0.010$.   
This correlation should be interpreted as evidence that the contribution 
of Fe to the total metallicity is depressed, rather than that alpha 
elements are enhanced, at high $\sigma$ (e.g., Worthey et al. 1992; 
Weiss, Peletier \& Matteucci 1995; Greggio 1997; Trager et al. 2000a).  

If the evolution in Mg and Fe is due to the same physical process, 
then one might have wondered if residuals from the Mg$b -\sigma$ 
relation are correlated with residuals from the 
$\langle{\rm Fe}\rangle - \sigma$ relation.  
This will be easier to address when the sample is larger.  
At the present time, we see no compelling evidence for such a 
correlation---we have not included a figure showing this explicitly.  
In addition, at any given redshift, galaxies which are richer in 
Mg$_2$ or $\langle{\rm Fe}\rangle$ than they should be (given 
their velocity dispersion), are neither more nor less likely to be 
richer in H$_\beta$ than expected---recent star formation is not 
correlated with metallicity.  

\subsection{Line-indices and color}\label{lindexc}
Because both the colors and the line indices are evolving, it is 
interesting to see if the evolution in color and in the indices is 
similar.  The line indices and color both correlate with $\sigma$, 
and we know how much the individual relations evolve, so we can estimate 
the evolution in the index--color relation as follows.  

Let $y_0 = s x_0 + c_0$ denote the mean relation between line index 
$y$ and color $x$ at $z=0$.  Because we know that line indices and color 
both correlate with $\sigma$, and we know how much the individual 
relations evolve, we can estimate the evolution in the index--color 
relation by setting 
$y(z) = y_0 + \Delta y = s x_0 + c_0 + \Delta y 
     = s x(z) + c_0 + \Delta y - s \Delta x$.
For $x=(g^*-r^*)$ and $y={\rm Mg}_2$ we must set $\Delta y = -0.015$, 
$\Delta x = -0.042$, and $s = 0.20/0.26$ (from Table~\ref{MLcmag} 
and Figure~\ref{fig:Mg2sigma}).  Thus the slope of Mg$_2$ versus 
$g^*-r^*$ color should have a slope of 0.77 and the zero-point is 
expected to evolve by 0.017 between the lowest and highest redshift 
bins in our sample.  A similar analysis for H$_\beta$ and $g^*-r^*$ 
suggests that slope is expected to be $-0.92$, and the zero point 
should evolve by $0.019$, whereas the slope of the 
$\log_{10}\langle{\rm Fe}\rangle$--color relation should be 0.42 with 
essentially no evolution.
(These estimates assume that the slopes of the individual relations do 
not evolve.  Bender et al. (1996) present some evidence that Mg$b$ at 
high $\sigma$ evolves less than at low $\sigma$, suggesting that the 
slope of the Mg$b-\sigma$ relation was steeper in the past. A comparison 
of the $\log_{10}\sigma=2.3$ and 2.4 bins in Figure~\ref{fig:lindices} 
is in approximate agreement with this.  Because these estimates are of 
the order of the error in the measurements, we have not worried about 
the effects of a change in slope---but with a larger sample, this will 
be important.)

\begin{figure}
\centering
\epsfxsize=1.1\hsize\epsffile{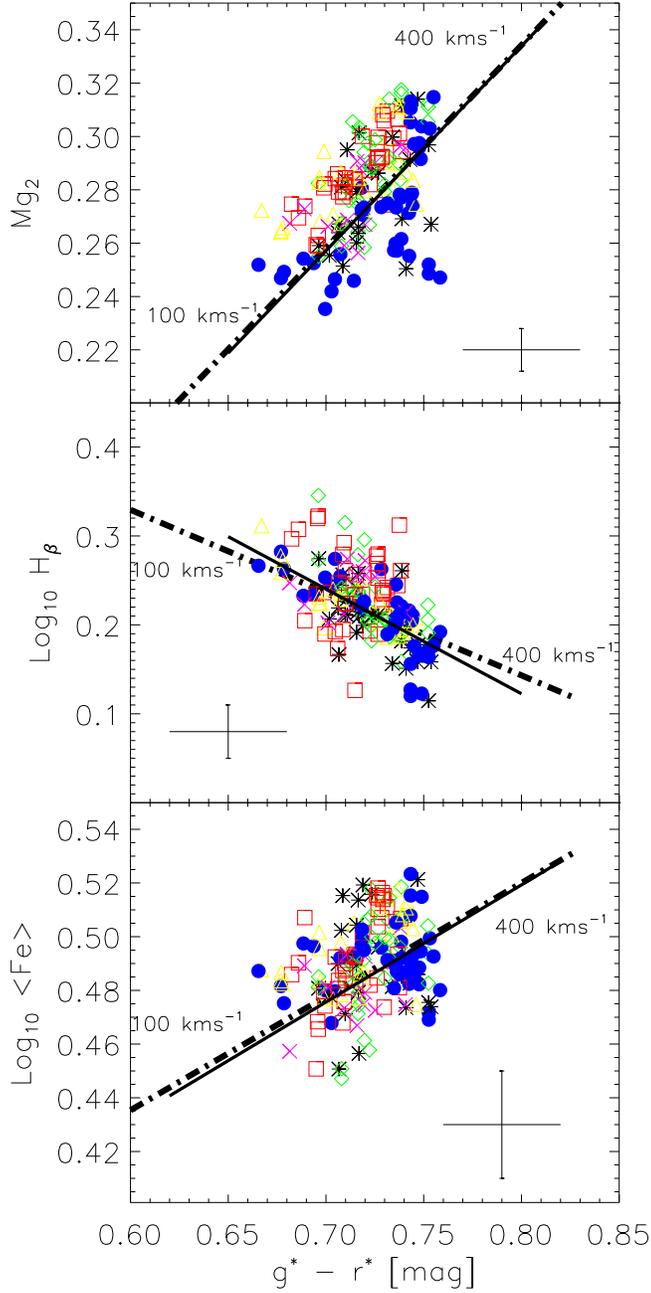}
\caption[]{Line indices versus color.  Dot-dashed lines show the slope 
one expects if there were no scatter around the mean color-$\sigma$ 
and lineindex--$\sigma$ relations, and solid lines show the linear 
relation which provides the best fit to the points.  Text along the 
each dot-dashed line indicates the typical value of the velocity 
dispersion at that location in line-index--color space.  
Crosses in the bottom of each panel show the typical uncertainties.  
The error in the color is supposed to represent the actual uncertainty 
in the color, rather than how well the mean color in each bin has been 
measured.  }
\label{lndxclr}
\end{figure}

To check the accuracy of these estimates Figure~\ref{lndxclr} shows 
plots of the line indices versus $g^*-r^*$ color.  (Recall that the 
line indices were computed from coadded spectra of galaxies which had 
the same $R_o$, $\sigma$ and $r^*$ luminosity.  The color here is the 
mean color of the galaxies in each of those bins.)  
The solid lines show best fits to the points contributed by the 
median redshift bins (triangles and diamonds).  
The dot-dashed lines in each panel show the slopes estimated above; 
they are not far off from the best-fit lines.  (The text in each panel 
indicates the typical velocity dispersion associated with different 
locations in index--color space.)  
The estimates of the evolution of the zero-point also appear to be 
reasonably accurate.  The higher redshift crosses in the Mg$_2$ panel 
appear to lie about 0.02~mags above the lower redshift stars; the 
difference between the low and high redshift populations is obvious.  
In contrast, the evolution in H$_\beta$ and color conspire so 
that there is little net offset between the low and high redshift 
populations (note that an offset of 0.02~mags in Mg$_2$ is much more 
obvious than an offset of 0.02 in H$_\beta$).  This suggests that 
the evolution in color and in $H_\beta$ are driven by the same process.  
And finally, there is little or no offset between the low and high 
redshift bins in $\langle{\rm Fe}\rangle$ and $g^*-r^*$ (bottom panel), 
as expected.  

We have also checked if the residuals from the index--color relations 
shown in Figure~\ref{lndxclr} correlate with local density:  they do not.  

\subsection{Comparison with stellar population models}\label{ssp}
The different line index--$\sigma$ and color--$\sigma$ relations are 
evolving.  Stellar population models can be used to study what the 
evolution we see implies.  

The predictions of single age stellar population models 
(e.g., Bruzual \& Charlot 1993; Worthey 1994; Vazdekis et al. 1996; 
Tantalo et al. 1998) are often summarized as plots of H$_\beta$ versus 
Mg$b$ (or Mg$_2$) and $\langle{\rm Fe}\rangle$.  The usual caveats 
noted by these authors about the limitations of these models, and the 
assumption that all the stars formed in a single burst, apply.  
In addition, comparison with data is complicated because the models 
assume that the ratio of $\alpha$-elements to Fe~peak elements in 
early-type galaxies is the same as in the Sun, whereas, in fact, it 
differs from the solar value by an amount which depends on velocity 
dispersion (e.g., Worthey et al. 1992; see bottom panel of 
Figure~\ref{fig:lindices}).  We use a simplified version of the 
method described by Trager et al. (2000a) to account for this.  

\begin{figure}
\centering
\epsfxsize=\hsize\epsffile{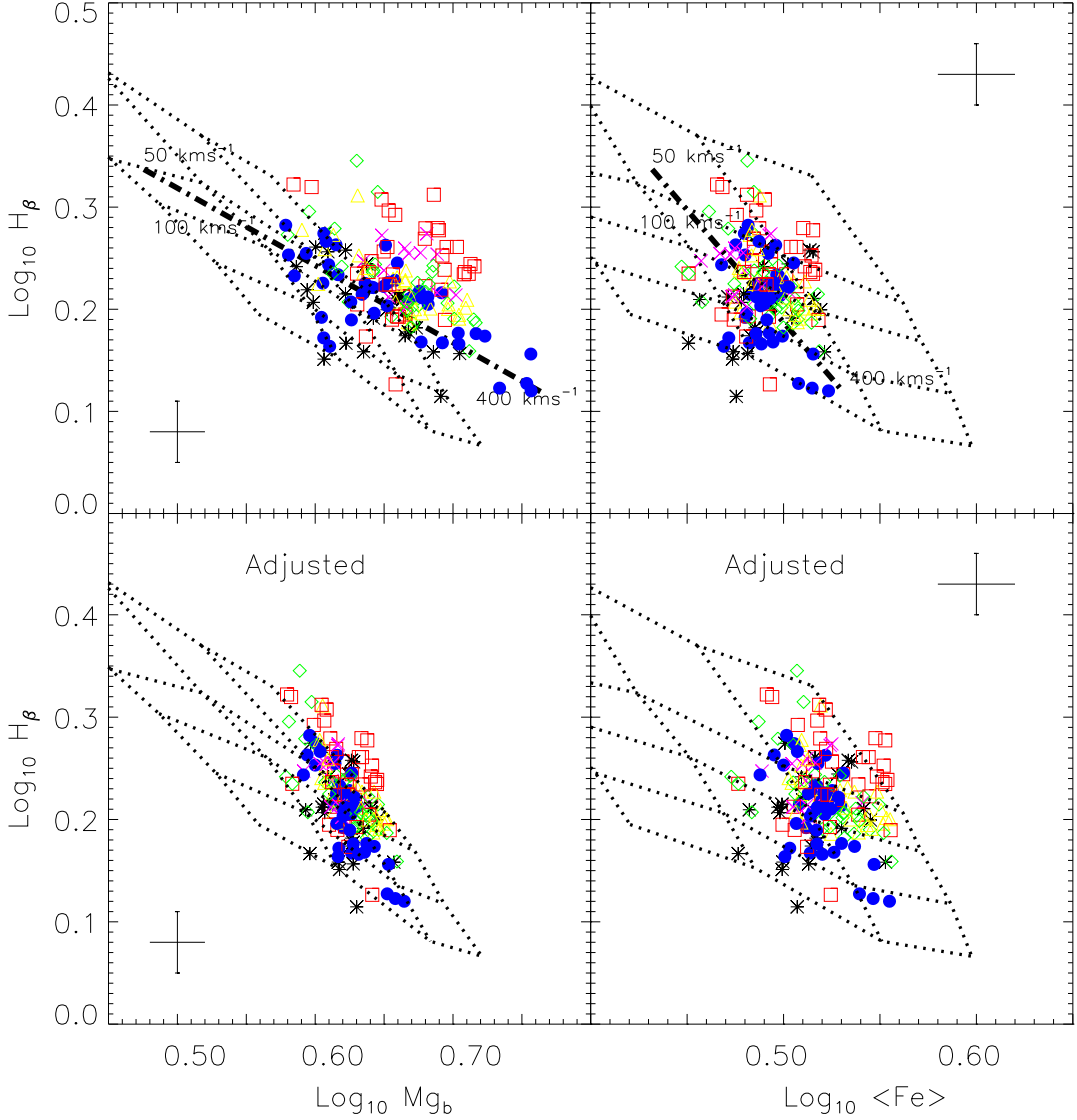}
\caption{H$_\beta$ versus Mg$b$ (left) and $\langle{\rm Fe}\rangle$ 
(right) for the coadded spectra in our sample.  Different symbols show 
different redshift bins;  the higher redshift population (squares 
and crosses) appears to show a larger range in H$_\beta$ compared 
to the low redshift population (stars and circles).  Cross in each 
panel shows the typical uncertainty on the measurements.  
Dotted grid shows a single age, solar abundance (i.e., [E/Fe] $=0$), 
stellar population model (from Worthey 1994); lines of constant age run 
approximately horizontally (top to bottom show ages of 2, 5, 8, 12 and 
17~Gyrs), lines of constant metallicity run approximately vertically 
(left to right show [Fe/H] $=-0.25$, 0, 0.25, 0.5).  
The two top panels provide different estimates of the age and metallicity, 
presumably because the [E/Fe] abundances in our data are different from 
solar.  In the bottom panels, this difference has been accounted for, 
and the age and metallicity estimates agree.  
Dot-dashed line in top panels shows what one 
expects if there is no scatter around the line index--$\sigma$ 
relations (solid lines in Figure~\ref{fig:lindices}); text shows the 
typical velocity dispersion associated with the location in 
`index--index' space.  }
\label{HbMgbFe}
\end{figure}

\begin{figure}
\centering
\epsfxsize=\hsize\epsffile{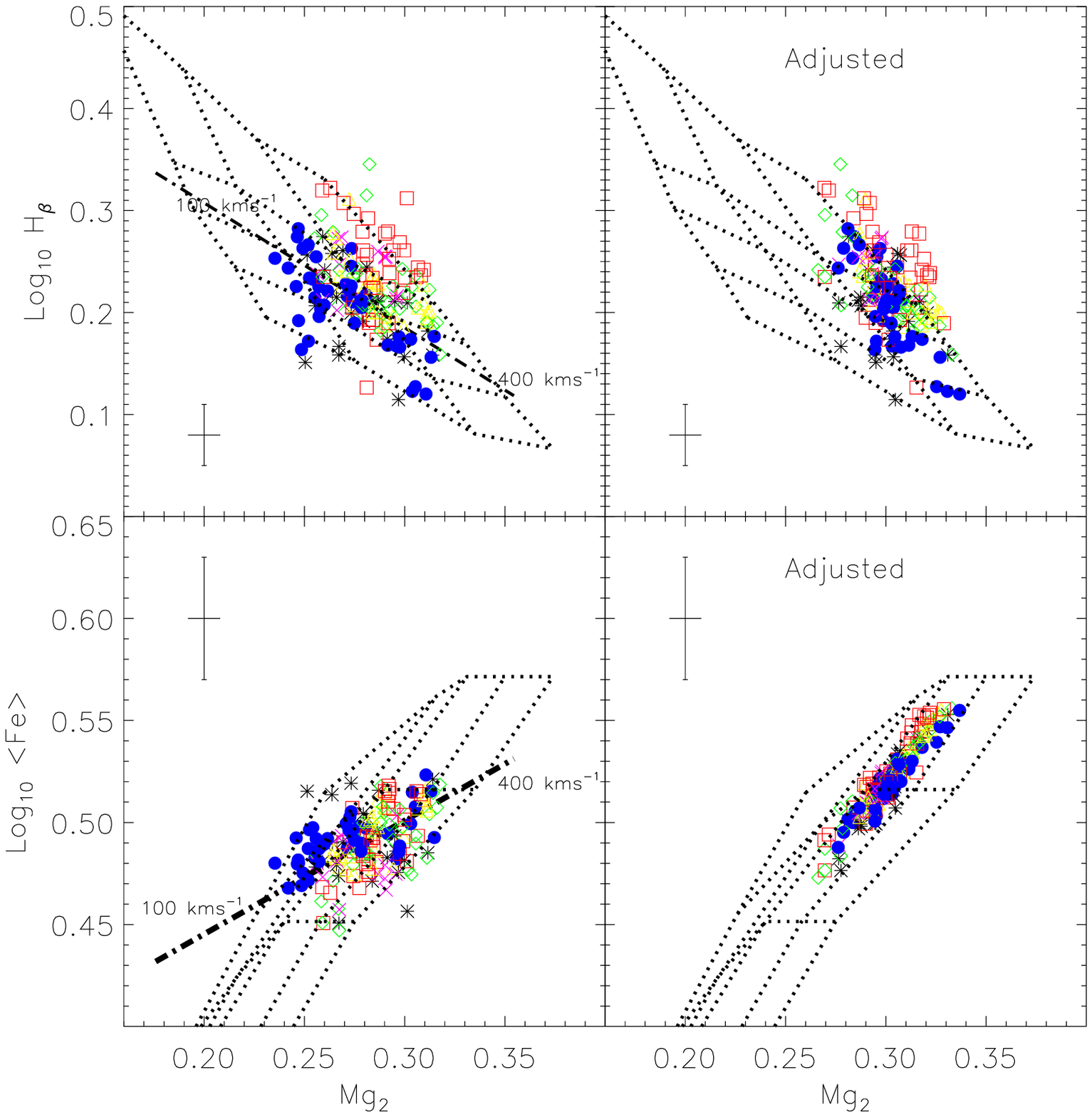}
\caption{Line indices H$_\beta$ and $\langle{\rm Fe}\rangle$ 
versus Mg$_2$ for the coadded spectra in our sample.  
Symbols (same as previous figure) show results for different redshift 
bins.  Dot-dashed line shows the relation one predicts if there were 
no scatter around the individual line index--$\sigma$ relations; text 
shows the typical velocity dispersion associated with the location in 
`index--index' space.  
Evolution is expected to move points upwards and to the left for 
H$_\beta$ versus Mg$_2$ (top panels), but downwards and left, and 
along the dot-dashed line in the case of $\langle{\rm Fe}\rangle$ and 
Mg$_2$ (bottom panels), although selection effects make these trends 
difficult to see.  Dotted grids show the same single stellar 
population model as in the previous figure (from Worthey 1994).
Age and metallicity estimates in the top panels are inconsistent with 
those in the bottom panels if solar abundance is assumed (left panels), 
but the estimates agree once differences in abundances have been accounted 
for (right panels). }
\label{fig:HbFeMg2}
\end{figure}

Figure~\ref{HbMgbFe} shows such a plot.  The dotted grids (top and bottom 
left are the same, as are top and bottom right) show a single age, solar 
abundance (i.e., [E/Fe] $=0$), stellar population model (from Worthey 1994); 
lines of constant age run approximately horizontally (top to bottom show 
ages of 2, 5, 8, 12 and 17~Gyrs), lines of constant metallicity run 
approximately vertically (left to right show [Fe/H] $=-0.25$, 0, 0.25, 0.5).  
Points in the panels on the top show the values of H$_\beta$ and Mg$b$ 
(left) and $\langle{\rm Fe}\rangle$ (right) for the coadded spectra in 
our sample.  Different symbols show different redshift bins;  the higher 
redshift population (squares and crosses) appears to show a larger range 
in H$_\beta$ compared to the low redshift population (stars and circles).  
Cross in each panel shows the typical uncertainty on the measurements.  

The heavy dot-dashed lines in the top panels show the relation between 
H$_\beta$ and Mg$b$ or $\langle{\rm Fe}\rangle$ one predicts if there 
were no scatter around the individual line index--$\sigma$ relations 
(shown as solid lines in Figure~\ref{fig:lindices}):  
${\rm H}_\beta\propto {\rm Mg}b^{-0.75}$ and 
${\rm H}_\beta\propto \langle {\rm Fe}\rangle^{-2.2}$.  
We have included them to help disentagle the evolution we saw in the 
individual line index--$\sigma$ relations from the effect of the 
magnitude limit.  The text shows the typical velocity dispersion 
associated with the location in `index--index' space.  
The evolution in the Mg$b -\sigma$, $\langle {\rm Fe}\rangle - \sigma$, 
and H$_\beta - \sigma$ relations suggest that the higher redshift sample 
should be displaced upwards and to the left, with a net shift in 
zero-point of about 0.03 and 0.02 in the upper left and right panels 
of Figure~\ref{HbMgbFe}.  (We estimate these shifts similarly to how 
we estimated the evolution in the line index--color relations.)  

Although the expected evolution is slightly smaller than the typical 
uncertainty in the measurements, the top two panels in 
Figure~\ref{HbMgbFe} do appear to show that the high redshift population 
(squares and crosses) is displaced slightly upwards.  
The shift to the left is not apparent, however, because of the selection 
effect:  evolution moves the high $\sigma$ objects of the high redshift 
sample onto the the lower $\sigma$ points of the low redshift sample, 
but the low $\sigma$ objects at higher redshift, which would lie clearly 
above and to the left, are not seen because of the selection effect.  
Note that the selection effect works so that evolution effects are 
suppressed, rather than enhanced in plots like Figure~\ref{HbMgbFe}; 
therefore, a simple measurement of evolution in the upper panels should 
be interpretted as a lower limit to the true value.  

The top two panels show that our sample spans a range of about 0.3 
or more in metallicity, and a large range of ages.  However, notice 
that the two panels provide different estimates of the mean ages and 
metallicities in our sample.  This is because the [E/Fe] abundances 
in our data are different from solar.  Trager et al. (2000a) describe 
how to correct for this.  Our measurement  errors in Mg and Fe are 
larger than theirs, so we have adopted the following simplified version 
of their prescription.  

Let [Z(H$_\beta,\langle{\rm Fe}\rangle$)/H] denote the estimate of 
the metallicity given by the top right panel of Figure~\ref{HbMgbFe}:  
this estimate uses the observed values of H$_\beta$ and 
$\langle{\rm Fe}\rangle$, and the Worthey (1994) solar abundance 
ratio models.  
Trager et al. (2000a) argue that non-solar abundances change the 
relation between [Fe/H] and the true metallicity [Z/H]:  
${\rm [Fe/H]} = {\rm [Z/H]} - A{\rm [E/Fe]}$, where $A\approx 0.93$.  
Trager et al. (2000b) argue that 
 ${\rm [E/H]}\approx 0.33\log_{10}\sigma - 0.58$, 
and that the relation is sufficiently tight that one can substitute 
$\sigma$ for [E/Fe].  Although we have not measured this relationship 
between [E/Fe] and velocity dispersion in our sample, we assume it is 
accurate.  This allows us to define a corrected metallicity 
[Z/H]$_{\rm corr}\approx$ 
[Z(H$_\beta,\langle{\rm Fe}\rangle$)/H] $+ 0.33\,A(\log_{10}\sigma - 0.58)$.
Trager et al. (2000a) also argue that correcting for nonsolar [E/Fe] 
makes a negligible change to H$_\beta$.  Therefore, we combine the measured 
value of H$_\beta$ with [Z/H]$_{\rm corr}$ to compute a corrected
age $\tau_{\rm corr}$.  We then use Worthey's model with these corrected 
ages and metallicities to obtain corrected values of Mg$b$ and 
$\langle{\rm Fe}\rangle$.  These are plotted in the bottom panels.  
By construction, the values of H$_\beta$ in all four panels are the 
same, and the age and metallicity estimates in the bottom two panels 
agree.  The differences between our top and bottom panels are similar 
to the differences between Figures~1 and~3 of Trager et al. (2000a),
suggesting that our simple approximate procedure is reasonably accurate.  

We apply the same correction procedure to plots of 
H$_\beta - {\rm Mg}_2$ and $\langle{\rm Fe}\rangle - {\rm Mg}_2$ 
in Figure~\ref{fig:HbFeMg2}.  
The dot-dashed lines in the panels on the left show 
$\log_{10}{\rm H}_\beta\propto -1.20\,{\rm Mg}_2$ and 
$\log_{10}\langle{\rm Fe}\rangle\propto 0.55\,{\rm Mg}_2$.  
Matteucci et al. (1998) report that a fit to a compilation of 
$\langle{\rm Fe}\rangle - {\rm Mg}_2$ data from various sources has 
slope 0.6.  The dot-dashed line does not appear to provide a good 
fit in the top panel, although this may be due to a combination of 
evolution and selection effects:  fitting the relation separately 
for different redshift bins and averaging the results yields a line 
which is more like the dot-dashed line.  

The expected evolution is upwards and to the left for 
H$_\beta - {\rm Mg}_2$ and down and left for 
$\langle{\rm Fe}\rangle - {\rm Mg}_2$, with net shifts in zero-points 
of $0.040$ and $-0.009$.  Thus, in the bottom panel, evolution moves 
points along the dot-dashed line.  As with the previous plot, the 
selection effect makes evolution difficult to see.  
The dotted grid shows the Worthey (1994) model for these relations.  
Comparison of the two bottom panels suggests that much of the scatter 
in the observed $\langle{\rm Fe}\rangle - {\rm Mg}_2$ relation is 
due to differences in enhancement ratios.  

We can now use the models to estimate the mean corrected ages and 
metallicities of the galaxies in our sample as a function of redshift.  
The mean metallicity is 0.33 and shows almost no evolution.  
The mean age in our lowest redshift bin (stars, median redshift 0.06) 
is 8~Gyrs, whereas it is 6~Gyrs in the highest redshift bin (crosses).  
The redshift difference corresponds to a time interval of 1.63~Gyr; 
if the population has evolved passively, this should equal the difference 
in ages from the stellar population models.  While the numbers are 
reasonably close, it is important to note that, because of the magnitude 
limit, our estimates of the typical age and metallicity at high redshift 
are biased towards high values, whereas our estimate of the evolution 
relative to the population at low redshift is probably biased low.  
Nevertheless, it is reassuring that this estimate of a formation time of 
8 or 9~Gyrs ago is close to that which we use to make our K-corrections.  

\begin{figure}
\centering
\epsfxsize=\hsize\epsffile{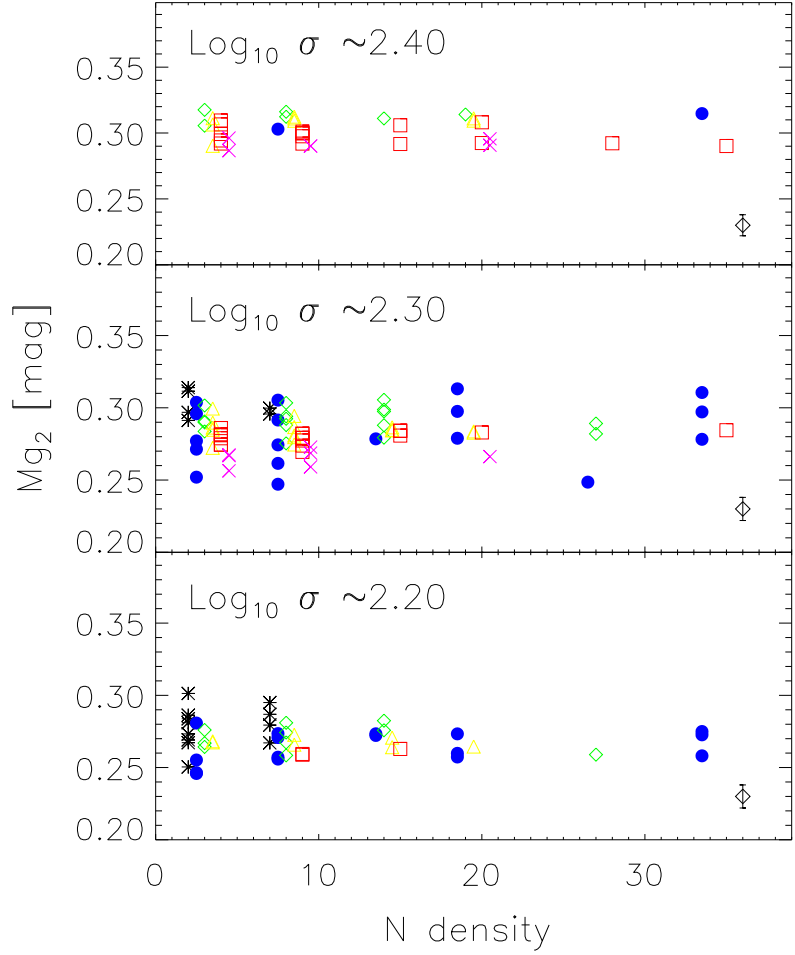}
\vspace{-2cm}
\caption{Mg$_2$--density relation for the galaxies in our sample.  
Symbols show the different redshift bins; higher redshift bins 
have been offset slightly to the right.  Symbol with bar in bottom 
right shows the typical uncertainty on the measurements. }
\label{densityMg2}
\end{figure}

\begin{figure}
\centering
\epsfxsize=\hsize\epsffile{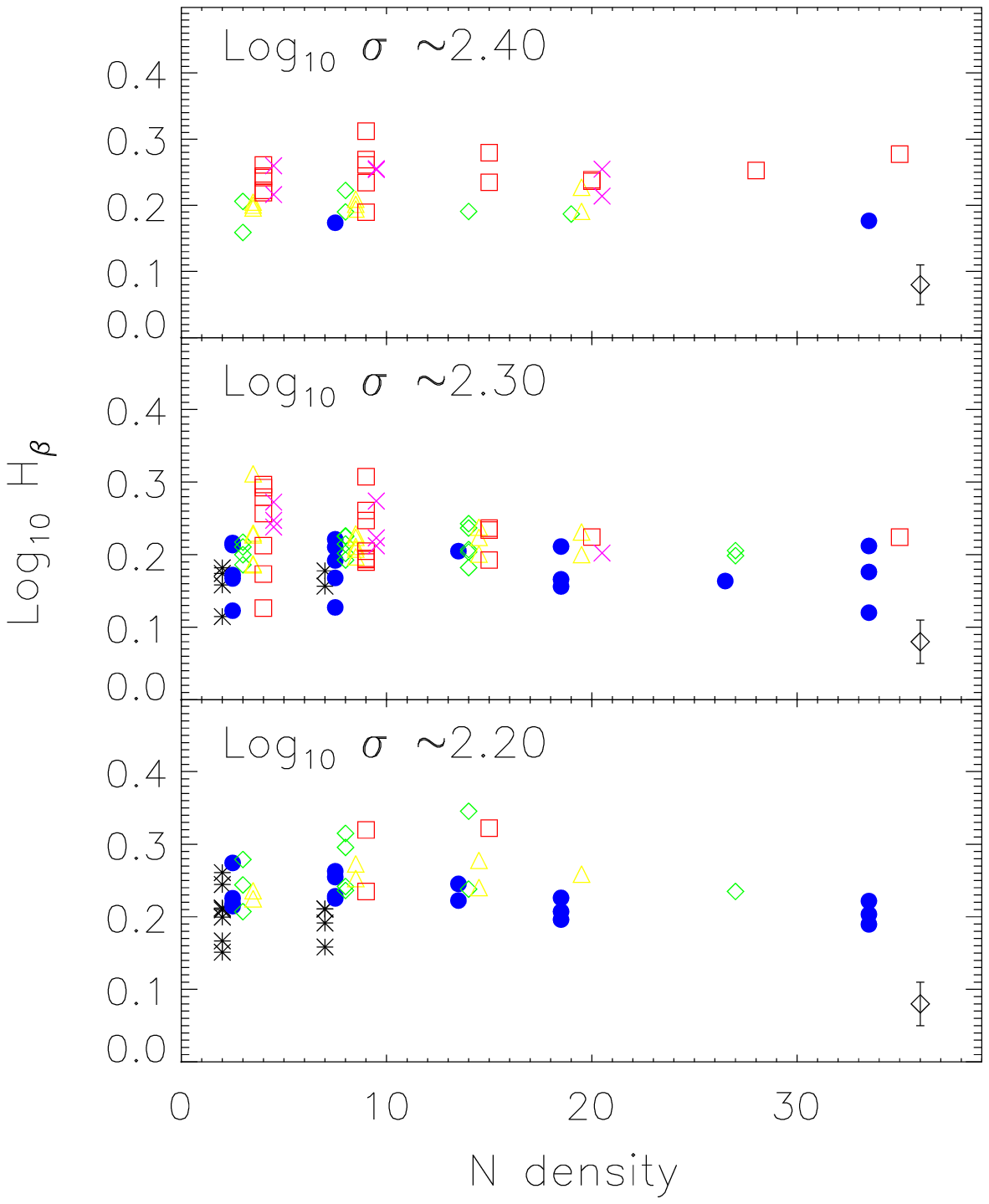}
\vspace{-2cm}
\caption{As for the previous figure, but for the H$_\beta$--density 
relation.  At fixed velocity dispersion, H$_\beta$ is slightly higher 
at high redshift, but there is no significant dependence on environment.}
\label{densityHb}
\end{figure}

\subsection{Dependence on environment}\label{environ}
We now turn to a study of how the coadded spectra depend on environment.  
Figures~\ref{densityMg2} and~\ref{densityHb} show the strength of Mg$_2$ 
and H$_\beta$ in a few small bins in $\sigma$, as a function of local 
density.  The different symbols in each density bin represent composite 
spectra from different redshifts---higher redshift bins have offset 
slightly to the right.  This allows us to separate the effects of 
evolution from those of environment.  
Figure~\ref{densityMg2} shows that, at fixed $\sigma$, Mg$_2$ 
decreases with redshift.  At any given redshift, the strength of Mg$_2$ 
is independent of local density.  (Our sample is not large enough 
to say with certainty if the evolution depends on environment.)  
A similar plot for H$_\beta$ also shows strong evolution with 
redshift, and no dependence on environment (Figure~\ref{densityHb}).  
Similar plots of $\langle{\rm Fe}\rangle$ and  [Mg$_2$/Fe] also 
show little if any dependence on redshift and no dependence on 
environment, so we have not shown them here.  

We caution that our definition of environment is limited, because 
it is defined by early-type galaxies only.  In addition, because 
we must divide our total sample up into bins in luminosity, size, 
radius, and redshift, and then by environment, the statistical 
significance of the results here would be greatly improved by 
increasing the sample size.  Analysis of environmental dependence 
using a larger sample is presented in Eisenstein et al. (2003).  

In conclusion, although we have evidence from the Fundamental Plane 
that early-type galaxies in dense regions are slightly different from 
their counterparts in less dense regions (Figure~9 in Paper~III; also 
see Figure~10 in Paper~I), these differences are sufficiently small 
that the strengths of spectral features are hardly affected 
(Figures~\ref{densityMg2} and~\ref{densityHb}).  
However, the coadded spectra provide strong evidence that the 
chemical composition of the population at low and high redshifts 
is different (Figures~\ref{fig:Mg2sigma}--\ref{fig:HbFeMg2}).  

\section{Discussion and conclusions}\label{discuss}
We have studied $\sim 9000$ early-type galaxies over the redshift 
range $0\le z\le 0.3$ using photometric (in the $g^*$, $r^*$, $i^*$ 
and $z^*$ bands) and spectroscopic observations.  
The colors of the galaxies in our sample are strongly correlated with 
velocity dispersion---redder galaxies have larger velocity dispersions 
(Section~\ref{cms}).  
The color--magnitude and color--size relations are a consequence of the 
fact that $M$ and $R_o$ also correlate with $\sigma$ (Figure~\ref{clrmagv}).  
The strength of the color--magnitude relation depends strongly on 
whether or not fixed apertures were used to define the colors, 
whereas the color$-\sigma$ relation appears to be less sensitive to 
these differences (Figures~\ref{fig:cmag} and~\ref{fig:csig}).  
At fixed velocity dispersion, the population at high redshift is bluer 
than that nearby (Figures~\ref{cmag3} and~\ref{csig3}), and the 
evolution in colour is significantly less than that of the 
luminosities (Table~\ref{MLcmag}).  
A larger sample, with well understood K-corrections, is required to 
quantify if galaxies in denser regions are slightly redder and more 
homogeneous or not (Figures~\ref{cmdensity} and~\ref{rmsdensity}).  

The SDSS spectra of individual galaxies do not have extremely high values 
of signal-to-noise (typically $S/N\sim 15$; cf. Figure~18 in Paper~I).  
However, the dataset is so large that we were able to study stellar 
population indicators (Mg$_2$, Mg$b$, $\langle{\rm Fe}\rangle$ and 
H$_\beta$) by co-adding the spectra of early-type galaxies which have 
similar luminosities, sizes, velocity dispersions, environments and 
redshifts to create composite with higher $S/N$ spectra.  The resulting 
library of 182 composite spectra, all of which have $S/N>50$, and many 
of which have $S/N>100$, covers a large range of velocity dispersions, 
sizes and luminosities.  It is available electronically.  

All the line indices correlate with velocity dispersion 
(Section~\ref{lindices}): Mg$_2\propto\sigma^{0.20}$, 
Mg$b\propto\sigma^{0.32}$, $\langle {\rm Fe}\rangle\propto\sigma^{0.11}$, 
and H$_\beta\propto\sigma^{-0.24}$.  These correlations are consistent 
with those in the literature, although the results from the literature 
were obtained from individual, as opposed to coadded, galaxy spectra.  
The coadded spectra show no signigicant dependence on environment.  
However, the spectra show clearly that, at fixed velocity dispersion, 
the high redshift population is stronger in H$_\beta$ and weaker in 
Mg and Fe than the population at lower redshifts 
(Figures~\ref{fig:Mg2sigma}--\ref{fig:HbFeMg2}).  
Line-indices also correlate with color:  a good approximation to 
these correlations is obtained by using the fact that line indices 
and color both correlate strongly with velocity dispersion, 
and ignoring the scatter.  

Single burst stellar population models (e.g., Worthey 1994; 
Vazdekis et al. 1996) allow one to translate the evolution in the 
spectral features into estimates of the ages and metallicities of the 
galaxies in our sample (e.g., Trager et al. 2000a,b).  In our sample, the 
$z\approx 0.05$ population appears to about 8~Gyrs old; 
the $z\approx 0.2$ population in our sample appears to be about 2~Gyrs 
younger; and the average metallicity appears to be similar in both 
populations.  The age difference is approximately consistent with the 
actual time difference in the $(\Omega_m,\Omega_\Lambda,h)=(0.3,0.7,0.7)$ 
world model we assumed throughout this paper, suggesting that the 
population is evolving passively.  Given a formation time, the single 
burst stellar population models also make predictions about how the 
luminosities and colors should evolve with redshift.  Our estimates of 
this evolution are also consistent with those of a population which 
formed the bulk of its stars 9~Gyrs ago.  

By the time the Sloan Digital Sky Survey is complete, the uncertainty 
in the K-corrections, which prevent us at the present time from making 
more precise quantitative statements about the evolution of the luminosities 
and colors, will be better understood.  In addition, the size of the sample 
will have increased by more than an order of magnitude.  This will allow 
us to provide a more quantitative study of the effects of environment than 
we are able to at the present time.  Most importantly, a larger sample 
will allow us to coadd spectra in finer bins; this will allow us to make 
maximum-likelihood estimates, rather than simple linear regression 
studies, of how features in the spectra correlate with other observables.  
This should also allow us to address the important issue of whether or 
not the most luminous galaxies in the population are evolving similarly 
to the faintest.

\vspace{1cm}

We would like to thank S. Charlot for making his stellar population 
synthesis predictions for the SDSS filters available to the 
collaboration.

Funding for the creation and distribution of the SDSS Archive has been 
provided by the Alfred P. Sloan Foundation, the Participating Institutions, 
the National Aeronautics and Space Administration, the National Science 
Foundation, the U.S. Department of Energy, the Japanese Monbukagakusho, 
and the Max Planck Society. The SDSS Web site is http://www.sdss.org/.

The SDSS is managed by the Astrophysical Research Consortium (ARC) for
the Participating Institutions. The Participating Institutions are The
University of Chicago, Fermilab, the Institute for Advanced Study, the
Japan Participation Group, The Johns Hopkins University, Los Alamos
National Laboratory, the Max-Planck-Institute for Astronomy (MPIA),
the Max-Planck-Institute for Astrophysics (MPA), New Mexico State
University, University of Pittsburgh, Princeton University, the United
States Naval Observatory, and the University of Washington.

{}

\end{document}